\documentclass[authoryear,preprint,review,10pt, 5p,times,twocolumn]{elsarticle}

\usepackage{graphicx}
\usepackage{amssymb}
\journal{Social Networks}
\providecommand{\e}[1]{\ensuremath{\times 10^{#1}}}
\usepackage[tight,normalsize,sf,SF]{subfigure}
\begin{document}

\begin{frontmatter}



\title{Predictors of short-term decay of cell phone contacts in a large scale communication network\tnoteref{tf1}\tnoteref{tf2}}
\tnotetext[tf1]{Research was sponsored in part by the Army Research Laboratory and was accomplished under Cooperative Agreement Number W911NF-09-2-0053 and in part by the National Science Foundation (NSF) Grant BCS-0826958. The views and conclusions contained in this document are those of the authors and should not be interpreted as representing the official policies, either expressed or implied, of the Army Research Laboratory or the U.S. Government. The U.S. Government is authorized to reproduce and distribute reprints for Government purposes notwithstanding any copyright notation hereon.}
\tnotetext[tf2]{We would like to thanks the \emph{Social Networks} editors and two anonymous reviewers for helpful comments and suggestions which served to significantly improve a previous version of this manuscript.}


%%\author[a]{}

%%\address{}
\author[ad1]{Troy Raeder}
\author[ad2]{Omar Lizardo\corref{corr1}}
\cortext[corr1]{Corresponding Author.}
\author[ad2]{David Hachen}
\author[ad1]{Nitesh V. Chawla}

\address[ad1]{Deparment of Computer Science and Engineering, College of Engineering, University of Notre Dame, 384 Fitzpatrick Hall, Notre Dame, IN, 46556}
\address[ad2]{Deparment of Sociology, University of Notre Dame, 810 Flanner Hall, Notre Dame, IN, 46556}

\begin{abstract}
Under what conditions is an edge present in a social network at time $t$ likely to decay or persist by some future time $t + {\Delta}t$?  Previous research addressing this issue suggests that the network range of the people involved in the edge, the extent to which the edge is embedded in a surrounding structure, and the age of the edge all play a role in edge decay.  This paper uses weighted data from a large-scale social network built from cell-phone calls in an 8-week period to determine the importance of edge weight for the decay/persistence process.  In particular, we study the relative predictive power of directed weight, embeddedness, newness, and range (measured as outdegree) with respect to edge decay and assess the effectiveness with which a simple decision tree and logistic regression classifier can accurately predict whether an edge that was active in one time period continues to be so in a future time period.  We find that directed edge weight, weighted reciprocity and time-dependent measures of edge longevity are highly predictive of whether we classify an edge as persistent or decayed, relative to the other types of factors at the dyad and neighborhood level.  
\end{abstract}


\end{frontmatter}

% \linenumbers

\section{Introduction}
Under what conditions are particular social connections more or less likely to dissolve over time? Most network analysts agree that the issue of the dynamic stability of social relationships embedded in networks is a fundamental one \citep{suitor_etal97, wellman_etal97, feld_etal07, bidart_degenne05}.  One obvious reason for the centrality of relationship dynamics is that essentially all of the classic behavioral theories in the network tradition---such as balance \citep{heider58, davis63} and exchange theory \citep{emerson72}---can be productively considered theories about the relative likelihood that some edges will persist and other edges will be dissolved \citep{hallinan78}.  For instance, classic balance-theoretic analyses of the dynamics of reciprocity suggest that the reason why we are more likely to observe tendencies toward reciprocity in human social networks is precisely because unreciprocated edges have a shorter lifespan---they are more likely to be dissolved by the unreciprocated party---and are thus weeded out of the network through a selection process \citep{hallinan78, tuma_hallinan78, hallinan_hutchins80, hallinan_williams87, bunt_etal99, van_duijn_etal03}.  A similar line of reasoning is behind Granovetter's \citeyearpar{granovetter73} influential ``strength of weak edges'' argument:  the reason why the intransitive ``forbidden triad'' is rare, is precisely because dyads embedded in fully-reciprocated triads are expected to be less likely to decay over-time \citep{davis67}---a proposition that has received some empirical confirmation by \citet{burt00}.  While much attention has been paid to the \emph{emergence} of transitivity in social networks through a process of meeting through an intermediary, it is clear that thinking dynamically about the \emph{persistence} of transitivity in social networks---through the selective dissolution of relationships \emph{not} embedded in triads---transforms this to a problem of accounting for the structural precursors of edge decay. This also implies that empirically ``bridges'' across transitive clusters should decay at a faster rate than other types of edges \citep{burt02}.  

In addition to these theoretical considerations, there are several substantive and practical motivations for the attempt to make progress in predicting edge persistent decay and persistence.  First, at the level of the whole network, edge decay may signal changing community structure \citep{Tantipathananandh_etal07}.  From an ego-centric perspective, if a given actor experiences high-levels of volatility and decay in her current relationships this  may indicate that he or she is moving between peer groups or undergoing a major life change \citep{suitor_keeton97, feld_etal07, bidart_lavenu05}. Second, relationships that are identified as likely to decay may under some circumstances (e.g. when there is a need to binarize a weighted matrix) be better thought of as ``false positives.''  In passively-collected behavioral data such as email and cell phone communications \citep{kossinets06, hidalgo_rodriguez08}---the source of data on which we rely in the analysis below---the notion of what exactly constitutes an edge is somewhat unclear.  Being able to predict edge decay may shed light on the circumstances under which an edge can be considered as ``real'' for the purposes of further analysis.

More recently, with the increasing availability of longitudinal social network data, the temporal evolution of social networks is beginning to receive increasing attention \citep{burt00, burt02, wellman_etal97, bidart_degenne05, bidart_lavenu05}. This has been aided by the recent development of actor-oriented, stochastic approaches for the analysis of longitudinal network data \citep{bunt_etal99, snijders05} which couple the evolution of micros-structures with agent-level attributes and behavioral outcomes (see \citet{snijders_etal10} for a recent review).  

However, in spite of its centrality for the main lines of theory in network analysis, the dynamics of link decay remains a relatively understudied phenomenon, especially at the level of behavioral observation.  In this respect, the main roadblock to a better understanding of the dynamics that drive patterns of decay of social edges in networks has been the relative paucity of large-scale, ecologically reliable data on social interactions \citep{eagle_etal08}.   The methodological and measurement issues associated with dynamic network data collected from informant reports on who they are connected to are well-known and well-documented, so there is no need extensively rehearse them here \citep{bernard_etal84, krackhardt87}.  These include: (1) systematic measurement error introduced by constraints on validity due to informant recall biases \citep{brewer00, brewer_webster00, marin04}, (2) measurement constraints  introduced due to reliance on so-called fixed-choice designs to accommodate for respondent's memory limitations and stamina \citep{feld_carter02, kossinets06}, and (3) validity limits introduced by data collection strategies that are limited (due to cost and the relative obtrusiveness of sociometric questionnaires) to small samples constrained to specific sites \citep{laumann_etal89} (4) limitations in the ability to measure the \emph{volume} and frequency of communicative activity that flows  through an edge, with most studies being relegated to using standard \emph{binary} networks in which links are thought to be either present or absent \citep{opsahl_panzarasa09, hammer85}.  

In the decay and formation dynamics of social relationships, these well-known limitations acquire renewed importance for three  reasons. First, as \emph{large-scale} (sometimes containing thousands---and in our case millions---of actors) data on human communication begins to accumulate, examining the extent to which standard analytic approaches can be used to account for empirical dynamics in this domain becomes a primary concern.  Second, when considering the issue of link decay, the problem of biases produced by memory limitations, artificial upper bounds on actor's degree produced by survey design strictures and the selective reporting of those contacts most subjectively (or objectively) important becomes an issue of substantive and methodological significance \citep{holland_leinhard73, kossinets06, kossinets_watts06}.  For while it is unlikely that persons will misreport being connected to those with whom they interact most often \citep{hammer85, freeman_etal87}, by selectively collecting data on ego's strongest edges, it is likely that survey-based methods may give undue consideration to precisely those links that are \emph{least likely} to decay. Finally, ignoring the fact that most real-world communication networks are not binary---each edge is instead ``weighted'' differently depending on the amount of communicative activity that flows through it \citep{barrat_etal04}---can impose artificial limits on our ability to predict which links are more likely to decay and which ones are more likely to remain.  In this paper we use behavioral network data on a large-scale sample of communicative interactions---obtained unobtrusively from cell-phone communication records---to study the dynamic and structural processes that govern link decay.  One advantage of the data that we use below is the fact that it consists of \emph{weighted} links based on dyadic communication frequency

It is of course not our intention to suggest that data obtained from cellular communication records are themselves devoid of bias or that previous research using self-report data do not constitute a solid foundation on which to build.  In fact, we rely on research and theory from such studies in the analysis that follows. Cellular communication data are certainly not a direct reflection of the underlying social network. Communicating by phone is only one out of a large menu of possible ways in which two persons may be connected; and in fact may persons can share strong connections without necessarily talking over the phone.  In addition just like informants may fail to mention their least important ties, rare-behavioral events (e.g. contacting somebody whom you only talk to once a year) will also be absent from observational data unless really long observation windows are used, thus producing a similar observational bias keyed to relative strength.

It is our contention however, that data obtained from spontaneous behavioral interactions will produce dynamical patterns that may be closer to those that govern the formation, sustenance and decay of human social relationships in ``the wild'' \citep{hammer85}.  As such, they are an important resource to establish the structural and dynamic properties of large-scale social networks.We already know that data of this type have high ecological validity, in that cell-phone mediated interaction accurately predicts face-to-face interaction and self-reported friendship as measured using traditional sociometric methods \citep{eagle_etal09}.  With penetration rates close to 100\% in industrialized countries such as the one from which these data were collected \citep{onnela_etal07}, cell-phone communications are also generally devoid of the socio-demographic biases that plagues studies that rely on modes of communication that have yet to achieve comparable levels of universal usage (such as email or chat).  \cite{onnela_etal07} examined basic topological properties of a cell-phone communication network similar to ours, and found it to display some basic signatures specific to social networks (e.g. small mean-path length, high-clustering, community structure, large-inequalities in connectivity across vertices, etc.).
  
This paper makes several contributions to the literature. First, on the substantive side, we incorporate insights and mechanisms from previous studies of network evolution to understand processes of link decay.  In addition we bring into consideration dyad-level process--such as degree of reciprocity--that have not yet been considered in studies of edge decay (mostly due to the fact that the data used are binary and not weighted).  On the methodological side, we introduce supervised learning techniques from the computer science literature for the study of social network evolution.  These techniques are appropriate for discovering patterns in data of the size and scope with which we are faced here (millions of persons and tens of millions of communication events), both extending and complementing the more traditional regression-based techniques that have been used to tackle this problem in the existing literature \citep[e.g][]{burt00, burt02}. Machine learning algorithms allow us to ascertain the relative importance of individual, dyadic and local-structural information in contributing to lowering or increasing the likelihood of link decay without incorporating strong assumptions about functional form---they are ``non-parametric'' in this respect---or homogeneity of effect sizes across the relevant feature space.  

The remainder of the paper is organized as follows: In the following section we briefly review previous research on edge decay in social networks.  In Section~\ref{sec:linkpred} we connect the substantive concern with identifying the factors that lead to link decay in the social networks with the largely methodological literature related to the link prediction problem in computer science and explain how we partially adapt these tools to the task at hand. In Section~\ref{sec:prevlit} we go on to review previous work on the dynamics of social relationships in large-scale networks.  In Section~\ref{sec:data} we describe the data on which we conducted this study and formally define each of the problems we consider.  Section~\ref{sec:statistics} describes basic topological and distributional features of our main predictors.  In Section~\ref{sec:correlation} we examine the correlation structure among the network features that we choose for the prediction task. In Section~\ref{sec:prediction} we present the results, identifying which network features are the strongest predictors of edge decay.  In Section~\ref{sec:model} we analyze the classifier's performance and explore their comparative fit. Finally in Section~\ref{sec:conclusions} we discuss the substantive implications of our results, draws conclusions, and lay out potential avenues for future research.

\section{Correlates of Edge Decay in Social Networks}
\label{sec:correlates}
A great deal of effort has gone into characterizing the growth of networks, either with high-level generative models (see \citep{chakrabarti2006gml} for a survey) or by analyzing the formation of individual links \citep{hays1984dam,marmaros2006ff}. Comparatively little work has been done on decay dynamics in large-scale networks with an already existing structure: the processes by which individual actors leave the network or individuals sever edges.  The most exemplary work on the issue of edge decay in social networks is that of \citet{burt00, burt02}, who studies the social networks of prominent bankers over time and analyzes the factors that contribute to the disappearance of edges.  Specifically, prominent bankers within an organization were asked, once a year for four years, to name other  bankers from the same organization with which they had had ``frequent and substantial business contact'' over the previous year.  Two main substantive conclusions emerge from this analysis:

\begin{enumerate}
\item {Several factors influence edge decay, including homophily (similarity between people), embeddedness (mutual acquaintances), status (e.g. network range), and experience.}
\item {Links exhibit a ``liability of newness'', meaning that newly-formed links decay more quickly than links that have existed for a long time.}
\end{enumerate}

These observations seem to lay out a framework for predicting link decay (and by implication, link persistence), and that is precisely the chief question of this paper: \textit{What are the vertex-level, dyad-level and local-structural features that can be used to most accurately predict edge decay?}  A formal statement of this research question gives rise to what we will call the \textit{decay prediction} problem: Given the activity within a social network in a time period $\tau_1$, how accurately can we predict whether a given edge will persist or decay in a following window $\tau_2$?  In what follows we evaluate the effectiveness of a machine learning solution to the decay prediction problem.  

\section{The Link Prediction Problem}
\label{sec:linkpred}
The problem of decay prediction is intimately related to the \textit{link prediction} problem.  There are several related but slightly different problems that are termed ``link prediction'' in the computer science literature.  The most related one, originally studied by
\citet{libennowell2007lpp} can be stated as follows: given the state of a network $G = (V, E)$ at time $t$, predict which \textit{new} edges will form between the vertices of $V$ in the time interval $\tau = (t, t + {\Delta}t)$.  See \citep{bilgic2007ccc,clauset2008hsa} for additional work
in this vein or \citep{getoor2005lms} for a survey.

Other authors \citep{kashima2006ppm} have formulated the problem
as a binary classification task on a static snapshot of the network, but this version of the problem is less related to the present effort simply because it is not longitudinal in nature.  Current research on link prediction in computer science focuses mostly on evaluating the raw predictive ability of different techniques, by either incorporating different vertex and edge attributes \citep{omadadhain2005par,omadadhain2005lpm,popescul2003srl} or the selecting different learning methods \citep{hasan2005lpu} in order to improve prediction performance.  Where we differ from this
work, apart from addressing a slightly different problem, is that we attempt to systematically characterize the attributes that lead to successful classification.  In other words, rather than being concerned simply with whether, or to what extent, our models succeed or fail, we attempt to characterize \textit{why} they are successful or unsuccessful by measuring the importance of different attributes and of weighted edge data to classification.

\section{Previous longitudinal research on large-scale networks}
\label{sec:prevlit}
Several authors have studied the evolution of large networks and identified characteristics that are important to the \textit{formation} of edges.   \citet{kossinets_watts06} studied the evolution of a University email network over time and the extent to which structural properties, such as triadic closure, and homophily contribute to the formation of new edges.  Of particular relevance to us, they find that edges that would close triads are more likely to form than edges that do not close triads, and that people who share common acquaintances are much more likely to form edges than people who don't.  Similarly, \citet{leskovec2008mes} study the evolution (by the arrival of vertices and the formation of edges) of four large online social networks and conclude, among other things, that triadic closure plays a very significant role in edge formation. Both of these factors are related to the notion of \textit{embeddedness} which we study in the context of edge decay, but neither of these authors consider edge decay at all.  \citet{marsili2004raf} develop a model for network evolution that allows for the disappearance of edges, but they do not validate the model on any real-world data.  As a result, the extent to which social networks fit the model is unclear and it does not shed any light on the mechanisms behind edge decay.  The effort that is closest to ours in principle is a paper by  \citet{hidalgo_rodriguez08}, which analyzes edge persistence and decay on a mobile phone network very similar to our own.  However, the analysis undertaken below differs critically from theirs both in methodology and primary focus. The aforementioned paper relies on a highly circumscribed set of well-established physical network statistics (i.e. degree, clustering coefficient) as well as reciprocity to explain decay. In what follows, we consider time-dependent properties of edges \citep{burt00} as well as features associated with interaction frequency (edge weight) \citep{marsden_campbell84, hammer85, barrat_etal04}.   

\begin{table*}[ht!]
	\centering
	\begin{tabular}{|l |c|}
		\hline
		\textbf{Statistic}&\textbf{Value}\\
		\hline
		\hline
		Average Clustering Coefficient ($d_i >= 2$):& 0.24\\
		Median Clustering Coefficient: &0.14\\
		Average Out-Degree: &4.2\\
		Median Out-Degree: &3\\
		Average Total Degree: &6.3\\
		Median Total Degree: &3\\
		Number of Vertices: & 4,833,408\\
		Number of Edges: &16,564,958\\
		\hline
	\end{tabular}
	\caption{ Basic graph-statistics of the cell-phone network.}
	\label{tbl:basic}
\end{table*}

\section{Data and features}
\label{sec:data}

\subsection{Data}
\label{subsec:data}

Our primary source of data in this study consists of information on millions of call records from  a large non-U.S. cell phone provider.  The data include, for each call, anonymized information about the caller and callee (i.e. a consistent index), along with a timestamp, duration, and the type of call (standard call, text message, voicemail call).   Our original dataset is composed primarily of phone calls and text messages.  In the empirical analysis that follows, however, we restrict ourselves to dyadic communications that take the \emph{exclusive form of a voice call} (we exclude text messages). We exclude all vertices with more that fifty neighbors, to ensure that only persons (and not auto-dialing robots) are represented in our data.  

Our final dataset consists of all in-network phone calls made over a 8-week period in 2008.  We restrict our attention to in-network calls (where both the caller and callee use our provider) because we only have information about calls initiated by our provider's customers.  That is to say, if $i$ is on the network but $j$ is not, we know if and when $i$ calls $j$, but not if and when $j$ calls $i$.  In order to accurately predict the decay of edges, we need to be able to capture the degree of \textit{reciprocity} in the relationship, meaning we need to be able to see if and when $j$ calls $i$ back.  Thus, we only examine edges where both $i$ and $j$ use our cell phone provider.

\subsection{Connectivity Criterion}
\label{subsec:connectivity}

Naturally, we represent this information as a directed social network, where the vertices are the individual subscribers. An edge exists from actor $i$ to actor $j$ if $i$ has at least one voice communication with $j$ during an initial window $\tau_1 = (t, t + {\Delta}t)$, which we define as $\tau_1=4^{weeks}$. Using this connectivity criterion, we identify  approximately 16.5\e{6} directed edges in the network (see Table~\ref{tbl:basic}).  Edges can be either bi-directional or directed arcs, depending upon whether $j$ made a call back to $i$ during  $\tau_1$.  Table~\ref{tbl:basic} shows some basic topological statistics of the observed graph. 

\subsection{Features}
\label{subsec:features}

Using the connectivity structure of the network constructed from the first four-weeks of data, we  extract a number of vertex-level, dyad-level and higher-order features based on the intuitions provided by previous research and theory on relationship dynamics \citep[e.g.][]{hallinan78, burt00, feld_etal07}, especially as they pertain to behavioral networks with weighted edges \citep{hammer85, barrat_etal04}.  These features are given in Table~\ref{tbl:features} and can be grouped into four categories or sets: vertex, dyadic, neighborhood, and temporal features.\footnote{We do not include any homophily-based features in this analysis, as we do not yet have reliable customer demographic information for the time period in question.}

\subsubsection{Vertex-level features}
The vertex-level features include the \emph{outdegrees} of $i$ and $j$ ($d_i$ and $d_j$), and the \emph{overall communicative activity} of each vertex ($c_i$ and $c_j$), that is the overall number of calls made by each member of the dyad during the 4-week time period, respectively.  

\subsubsection{Dyad-level features}
The dyadic level features include the directed arc strength, i.e. the counts of the number of voice calls made by $i$ to $j$ ($c_{ij}$), and the number of calls made by $j$ to $i$, $c_{ij}$.  We also compute normalized versions of arc strength ($p_{ij}$ and $p_{ij}$) which are simply the proportion of all calls made by an agent that go to that neighbor, where $p_{ij}=c_{ij}/c_i$.  

\subsubsection{Neighborhood-level features}
The neighborhood-level features include (1) the number of common neighbors between $i$ and $j$ ($cn$), (2) directional versions of the number of common neightbors ($in$ and $jn$) which indicate the number of $i$'s (or $j$'s) neighbors that called $j$ (or $i$), and (3)  second order embededness features ($injn$ and $jnin$) which measure the number of edges among $i$ and $j$'s neighbors.  $injn$ does this by counting as an edge calls made from one of $i$'s neighbors to one of $j$' neighbors, while $jnin$ considers an edge as existing when one of $j$' neighbors calls one of $i$'s neighbors. 

\subsubsection{Temporal features}
Finally we look at two features related to the (observed) temporal evolution of dyadic communicative behavior:  $fdate$ captures \emph{edge newness} as indicated by the time of the first call from $i$ to $j$ during our temporal window $\tau$; $fdate$ marks far into our time window, $\tau$, we \emph{first} observe a call from $i$ to$j$.  Higher values indicate newer edges, while smaller values indicate older edges.  The second temporal feature, $edate$, captures the \emph{edge freshness} as indicated by the time of the last call made by $i$ to $j$, given that the edge has already been observed to exist.  Higher values indicate that the edge was active in the more recent past, while smaller values indicate that the edge has been inactive for a longer period of time.  

To the best of our knowledge, prior work has not considered the freshness of an edge as a predictor of persistence/decay, though edge newness has been seen as as an important predictor of short-term decay via Burt's isolation of the phenomenon of the ``liability of newness'' of social ties \citep{burt97, burt00}.  We believe that the freshness of an edge could be an important predictor as it indicates how  \textit{current} the edge is and we  expect that more current edges are more relevant in the immediate future.  If persistence or decay are partly a markov process with a relatively short memory, then edge-freshness should emerge as an important predictive factor.

\subsection{Edge-decay and edge-persistence criterion}
\label{subsec:class_criterion}
\begin{figure*}[ht]
\centering
       \includegraphics[width=4.5in]{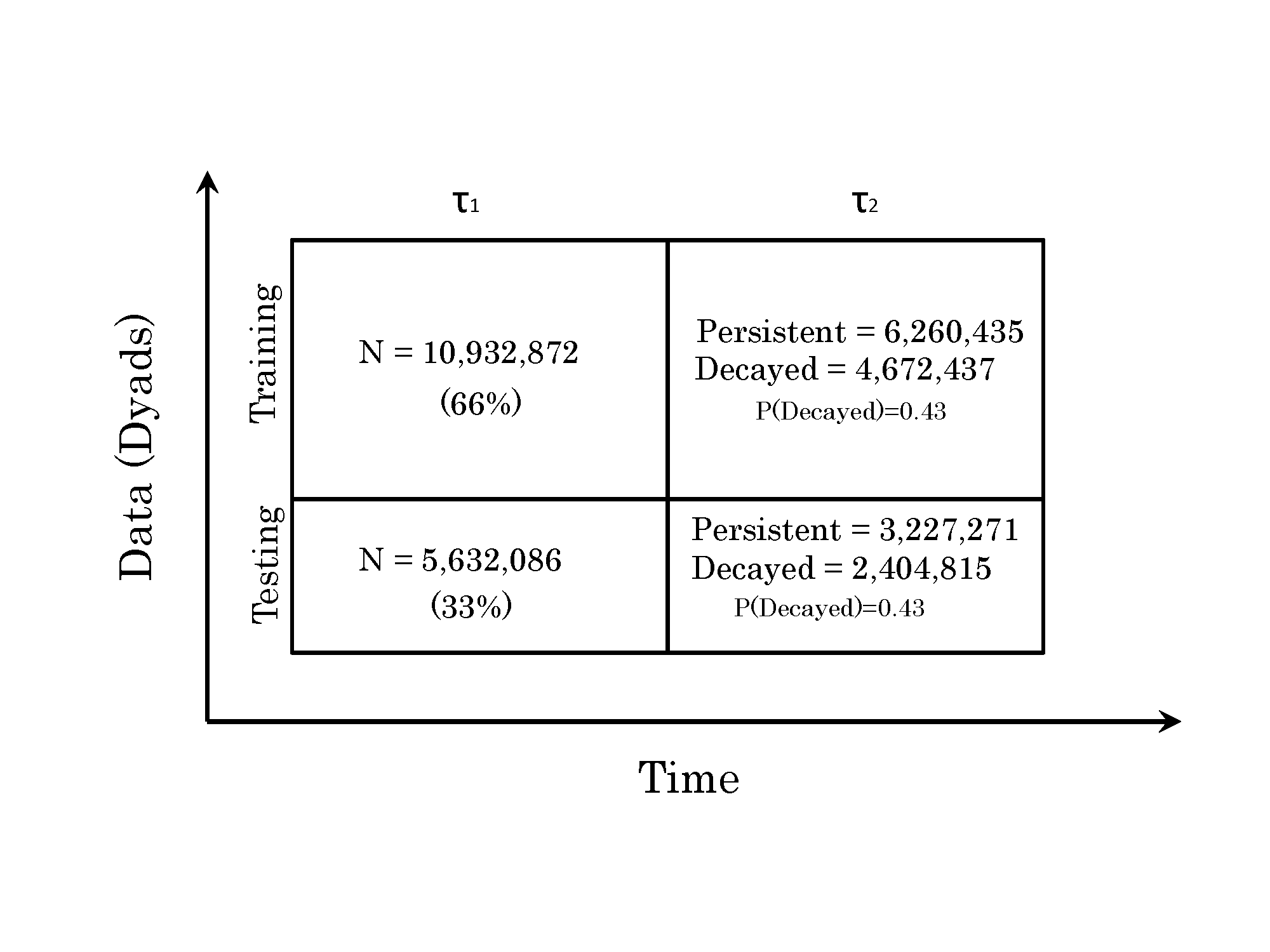}      	
	\caption{Diagram of the data-splitting procedure used to generate basic features and to determine dyadic class-membership in the analysis below.}
	\label{fig:method}
\end{figure*}

We use these features to build a model for predicting whether edges fall into two disjunctive classes:  \emph{persistent} or \emph{decayed}.  For the purposes of this analysis an edge is said to persist if it is observed to exist in the time period $\tau_2 = (t + {\Delta}t, t + {\Delta}t + {\Delta}t')$  given that it was observed in the previous time period $\tau_1 = (t, t + {\Delta}t)$ using the same connectivity criterion outlined in section~\ref{subsec:connectivity} above.  Conversely, an edge is said to have decayed if it was observed to exist in $\tau_1 = (t, t + {\Delta}t)$ but it can longer be detected in $\tau_2 = (t + {\Delta}t, t + {\Delta}t + {\Delta}t')$ using the same operational criterion.\footnote{We believe that a 4 week period is a long enough time window for determining edge persistence/decay at least in the short to medium term.  While technically an edge could be inactive during this period and reappear afterwards, all indications are that very few edges are like this, and those that are are very weak and fleeting.  While we could have lengthened the  $\tau_2$ time period, this would have meant shortening the $\tau_1$ period, but doing so would have affected our estimates of edge features that we use in the analysis.   Given that we have a total time window of $\tau_1+\tau_2=8^{weeks}$, we decided that the best strategy is to divide the period in half and define decay as the non-occurrence of voice calls between $i$ and $j$ in the second time period.}   Note that the observation and criterion periods are evenly divided such that $\tau_1=\tau_2=4^{weeks}$ (see Figure~\ref{fig:method}).

\section{Machine-Learning models for the edge-classification problem}
\label{sec:machine}

Having obtained a set of structural features from the network built from the information observed in $\tau_1$, our final task is to build a model that will allows us to most effectively assign each edge to either the persistent or decayed class using the criterion outlined in section~\ref{subsec:class_criterion} above.  Given the large scale of our communication network, we turn to methods from data-mining and machine-learning to accomplish this task.  

We proceed by arranging the available data as a set of \textit{instances} or \textit{examples}, each of which is observed to belong to a given class, which in our case is either persistent or decayed. As we noted above, associated with each instance is a set of \textit{features} or \textit{attributes}.  The task is to build a generalizable model from the available data.  In our case, since our class takes only the value 0 (decayed) or 1 (persistent), we need to derive a function $\mathbf{F} : \mathbf{x} \rightarrow \{0, 1\}$ which predicts (with some ascertainable accuracy) the class of an attribute given a vector of features $\mathbf{X}$.  

After building the model, we need to validate its effectiveness on a set of instances that are different from those used to build the model.  Typically this is done by dividing the available data into two disjoint subsets (the horizontal line in Figure~\ref{fig:method}).   The first subset, called the \textit{training data}, is used to build the model.  Once the model is built, we use it to predict the class of each instance in the test data.  The effectiveness of the model, then, depends on its accuracy (or some other measure of performance) on the test data.   As shown in Figure~\ref{fig:method}, the data are split \emph{within} each time period ($\tau_1, \tau_2$) into training and test subsets. In the analysis reported below, we randomly designated 2/3 of the original examples in the data in the first period to the training set and used the remaining 1/3 of the data as the testing set.  The figure also shows the number of dyads in testing and training set that ended up in the decayed class (about 43\%).  

\begin{figure*}[ht]
	\centering
	\subfigure[ ]{
	\includegraphics[width=3in]{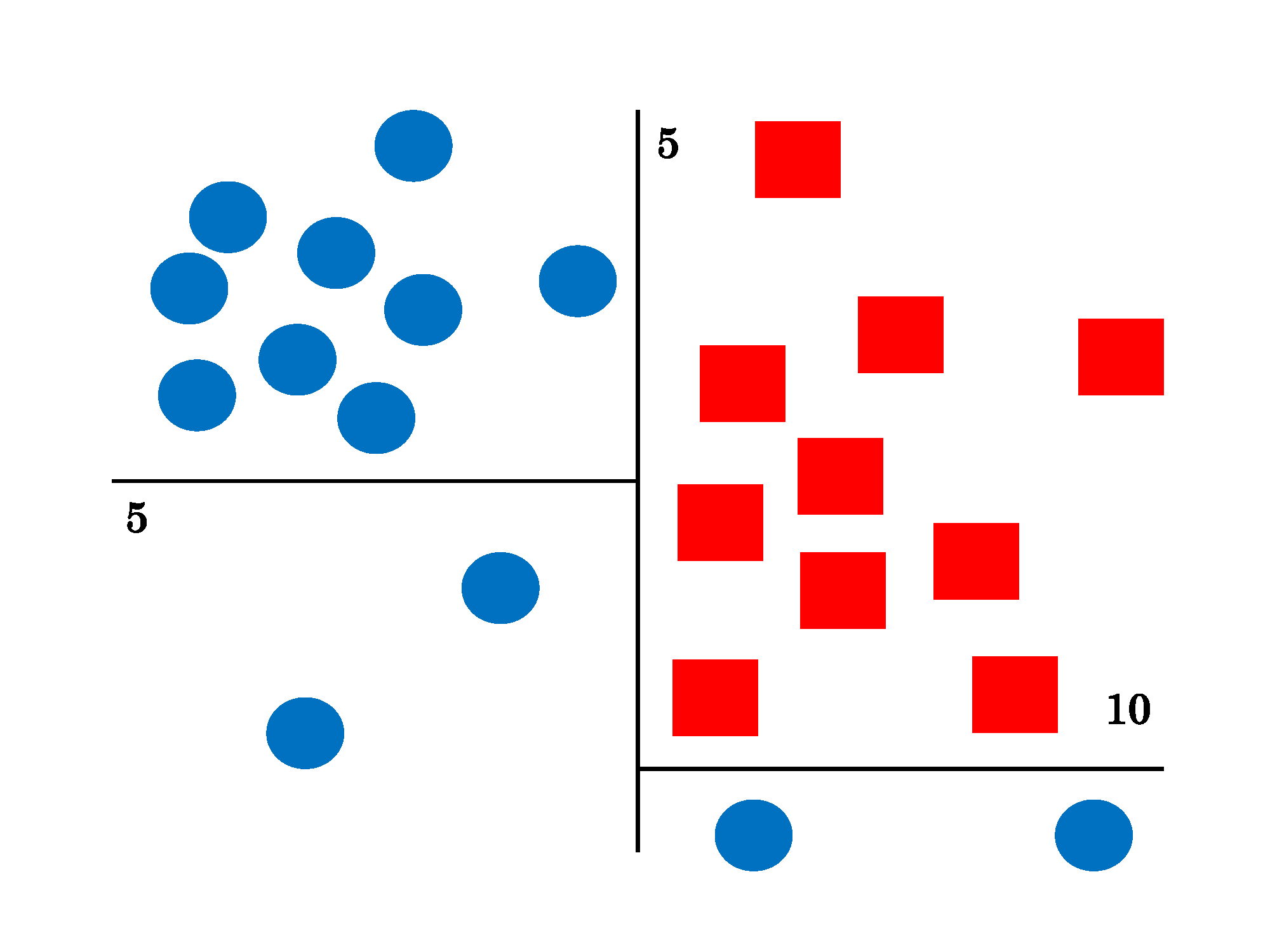}
	        \label{fig:toy-data}
		}
	\centering
	\subfigure[ ]{
	\includegraphics[width=3in]{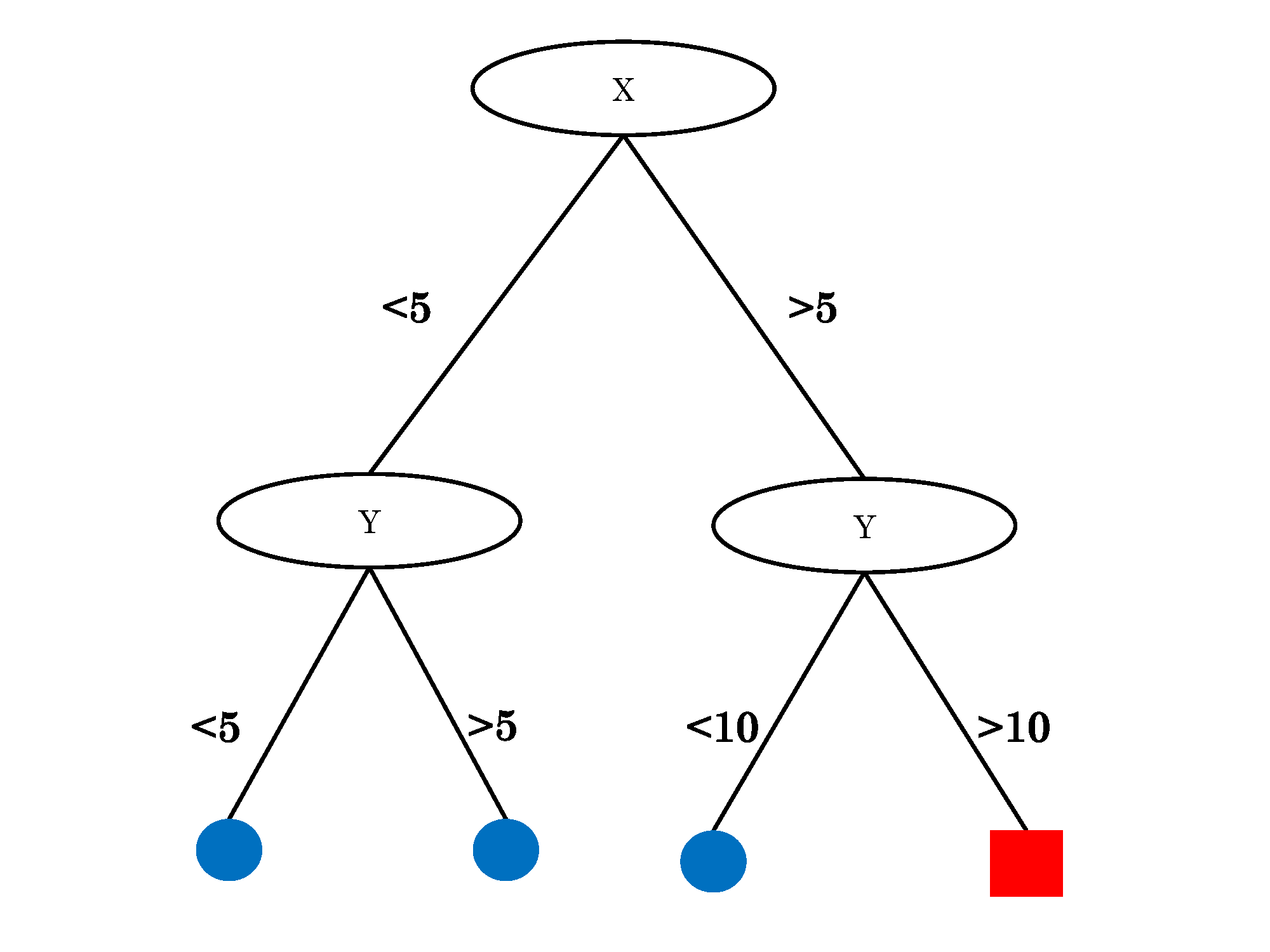}
	        \label{fig:toy-tree}
		}
	\caption{(a) Toy classification dataset.  (b) Resulting decision tree}
	\label{fig:example}
\end{figure*}

\subsection{The decision-tree classifier}
\label{subsec:tree}
There are a number of potential models available for evaluating the effectiveness of our chosen features at predicting edge decay.  Perhaps the simplest of these, which was used in Burt's analysis \citep{burt00}, is simple regression: plotting each feature as the independent variable against the probability of decay.  Such an approach has the distinct advantage that it is easy to interpret.  Regression is of course one of many classification tools available and also has the advantage of relative ease of interpretation.  In what follows we present results obtained from both a logistic regression classifier (which provides easily interpretable output that can be compared with previous research on the subject) and a decision-tree classifier which is an approach that has not been used very often in the analysis of Social Networks.\footnote{An important consideration with machine learning methods, as with statistical methods, is the choice of model to which we attempt to fit the data.  A nearly boundless series of models has been developed in the literature \citep{witten_frank05}, and a discussion of the merits of each is well beyond the scope of this paper.  Instead, we will discuss one of the models we chose (a \textit{decision tree} model called C4.5 \citep{quinlan1993cpm}) and its relative strengths for the problem at hand.} While relatively unfamiliar in the analysis of social networks, the decision tree is the most well-known and well-researched method in data-mining and provides output that is easily translatable into a set of disjoint ``rules'' for (probabilistically) assigning different cases to one of the two outcomes.  In our case we are interested in what combination of features maximize either edge decay or edge persistence. Because readers may not be wholly familiar with the decision-tree classifier approach, we provide a brief introduction to the basics of the approach before presenting the results.  We presume that readers are familiar with the basics of logistic regression so we will not discuss it in detail.

A decision tree classifies examples with a hierarchical set of rules.  A decision tree model is built by recursively dividing the feature space into \textit{purer} (more discriminating) subspaces along splits that are parallel to the feature axes.  A very simple example is shown in Figure~\ref{fig:example}.  Given the task of classifying unknown points as either blue circles or red squares (Figure~\ref{fig:toy-data}), a decision tree trained on the data points shown will produce a series of splits along the two dimensions in the figure ($x$ and $y$).  Generally, the first split is along whichever attribute is deemed to be the best separator of the classes according to some measure.  Our implementation of C4.5 determines the ``best'' split using information gain (which we will formally define in Section~\ref{sec:prediction} below).  Our hypothetical decision tree makes its first split along the line down the middle of the figure ($x = 5$)

\begin{figure*}[ht!]
\centering
\subfigure[\scriptsize  Outdegree of $i$ ] {
        \includegraphics[width=1.5in,  height=1.75in]{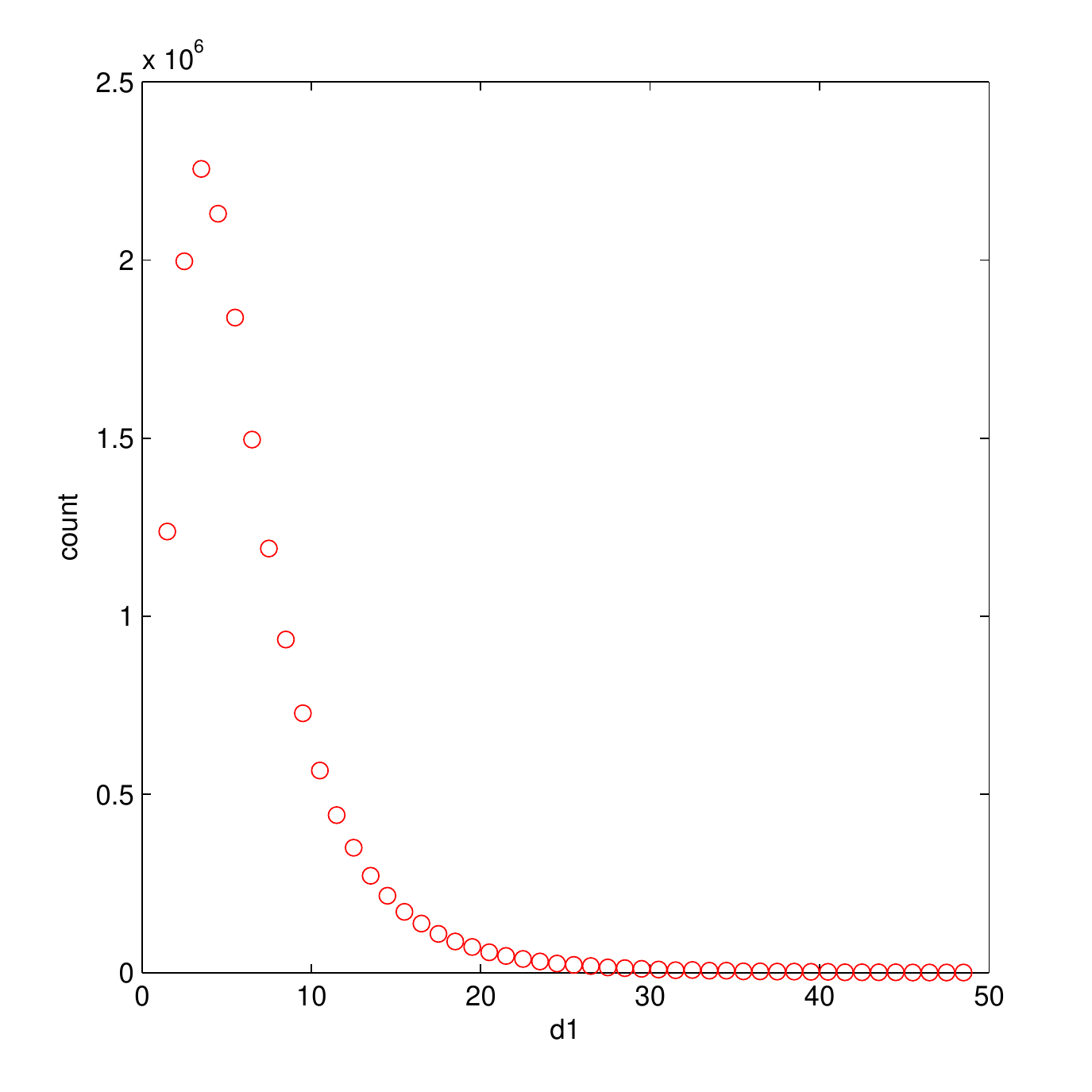}
        \label{ }
	}
\subfigure[\scriptsize Outdegree of $j$  ] {
        \includegraphics[width=1.5in,  height=1.75in]{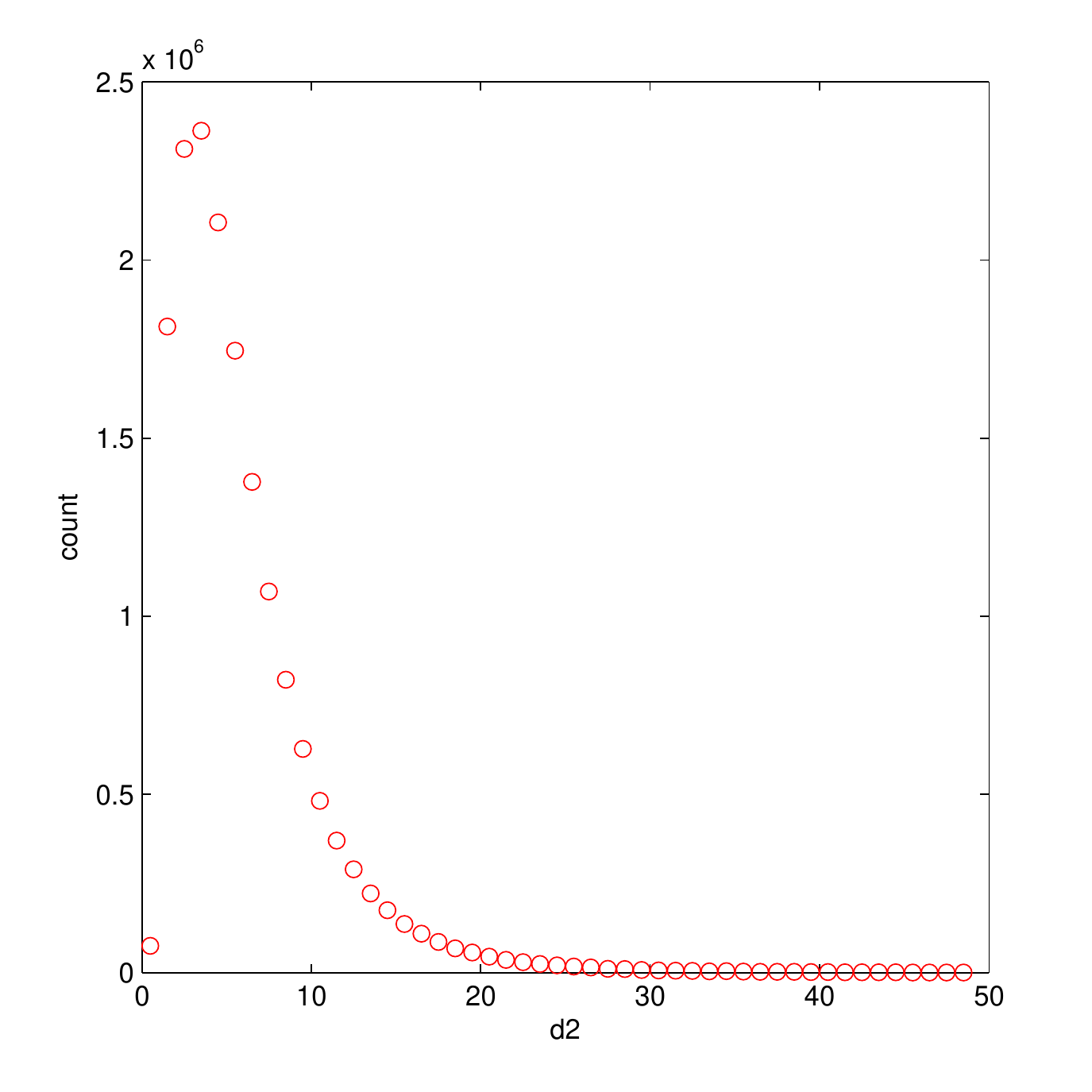}
        \label{ }
	}
\subfigure[\scriptsize N.  of calls made by $i$ ] {
        \includegraphics[width=1.5in,  height=1.75in]{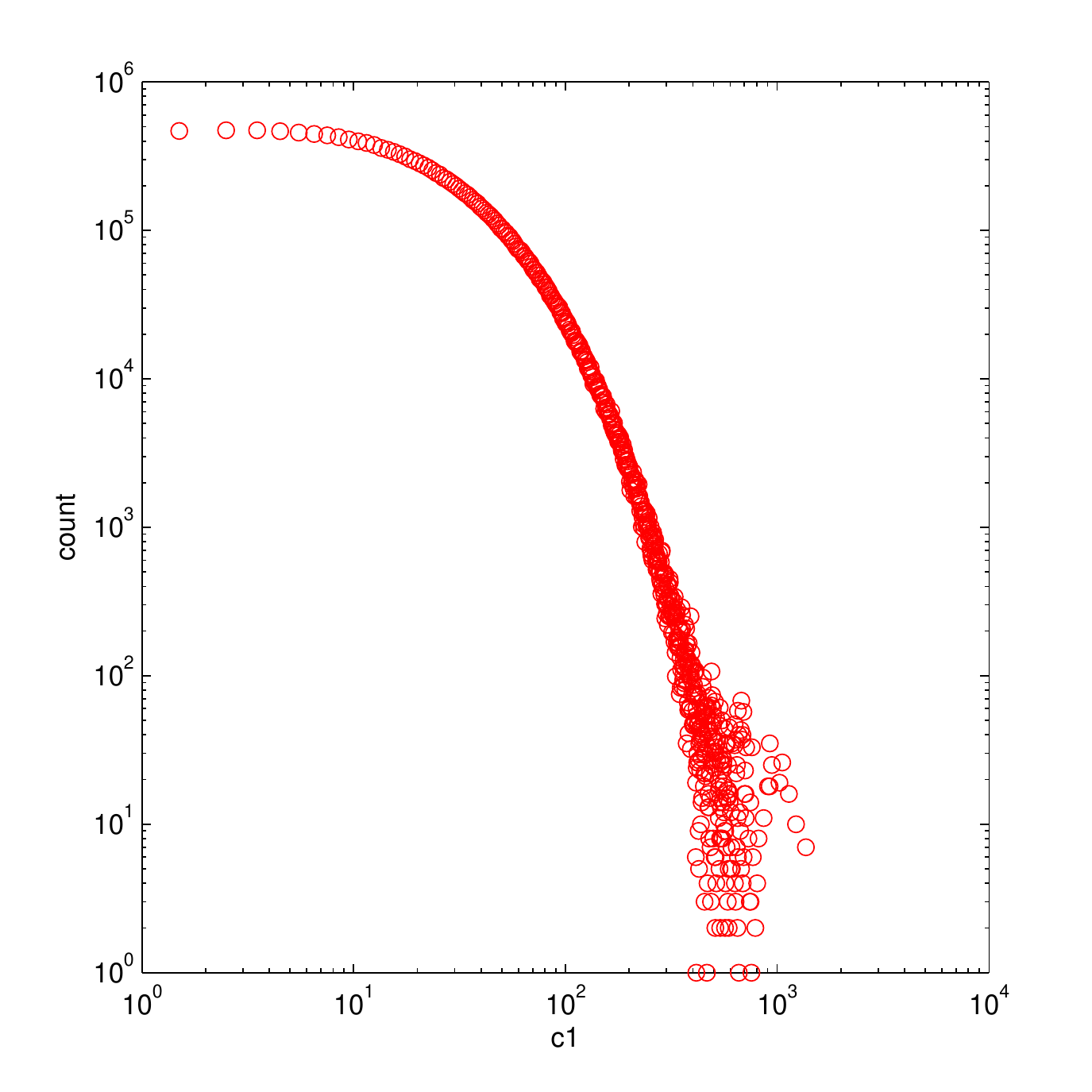}
        \label{ }
	}
\subfigure[\scriptsize N.  of calls made by $j$ ] {
        \includegraphics[width=1.5in,  height=1.75in]{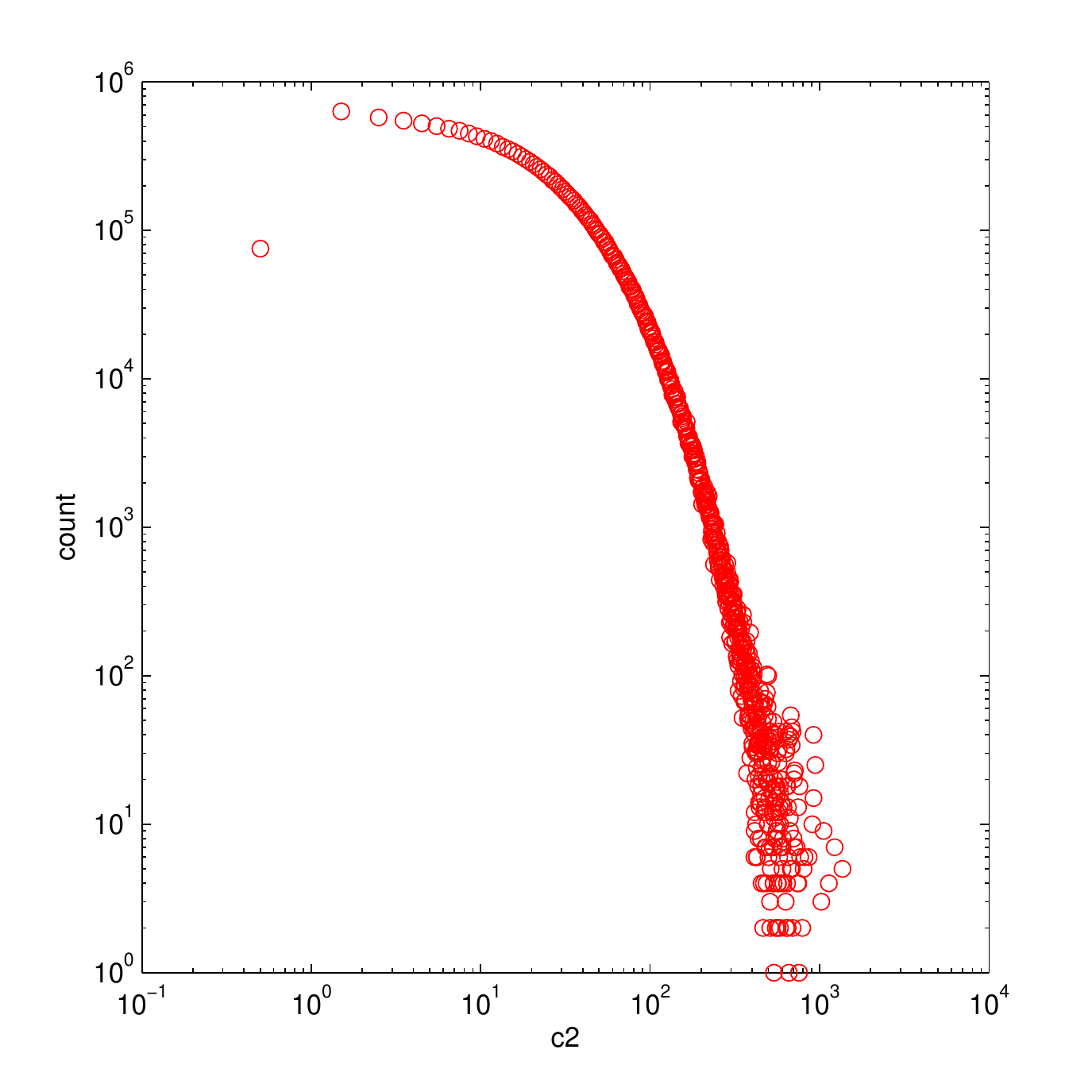}
        \label{ }
	}
\subfigure[\scriptsize  Calls from $i$ to $j$] {
        \includegraphics[width=1.5in,  height=1.75in]{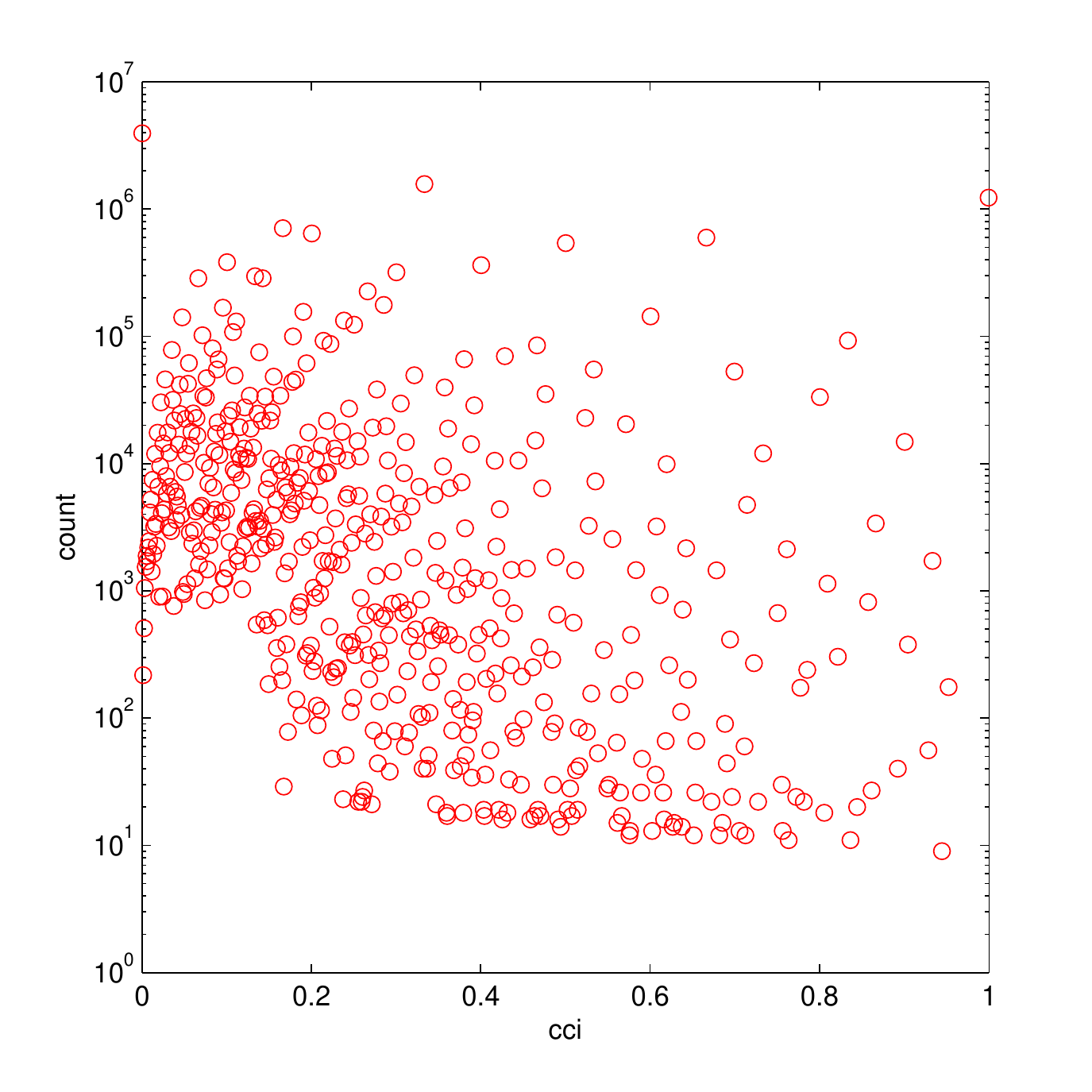}
        \label{ }
	}
\subfigure[\scriptsize  Calls from $j$ to $i$] {
        \includegraphics[width=1.5in,  height=1.75in]{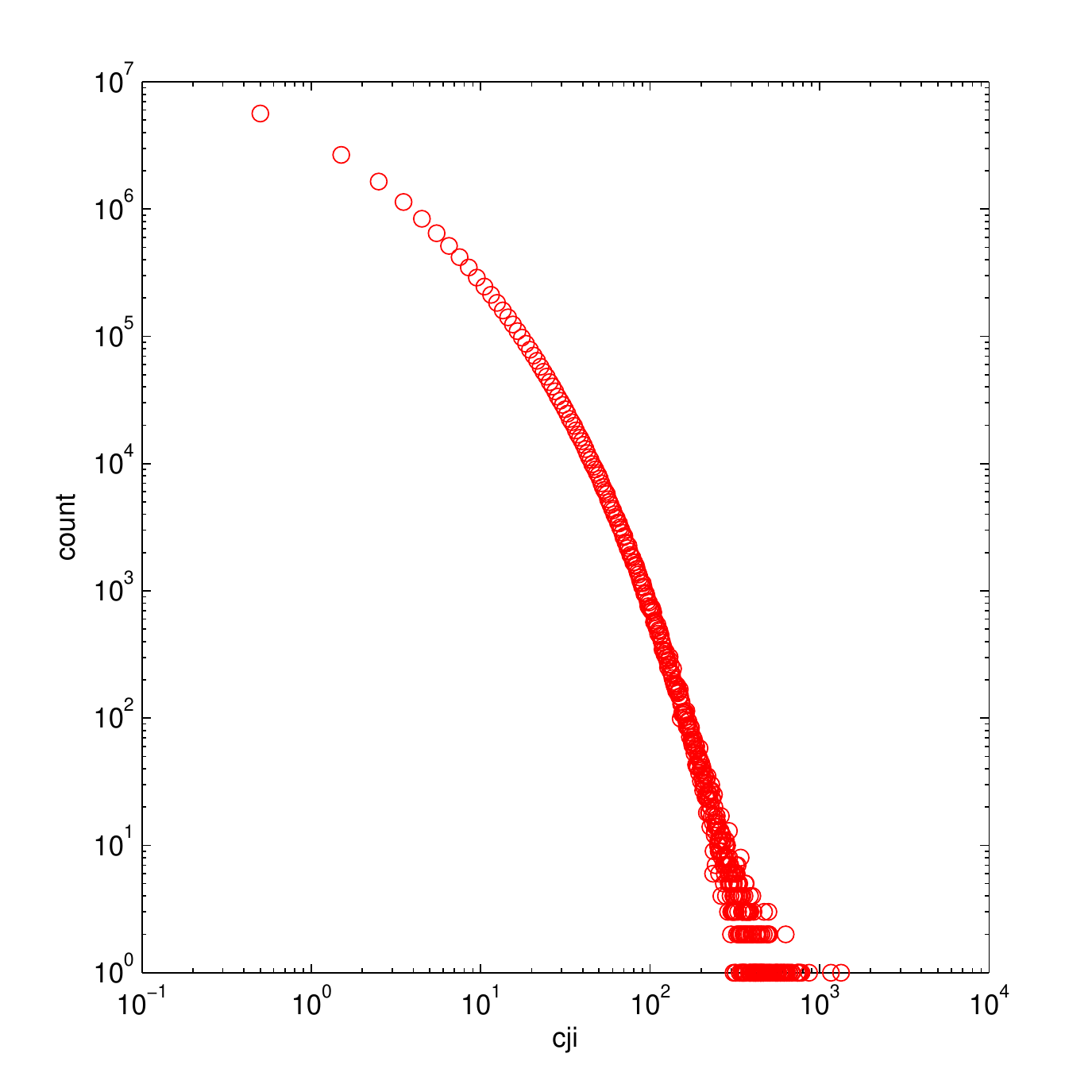}
        \label{ }
	}
\subfigure[\scriptsize  Prop. of $i \rightarrow j$ calls ] {
        \includegraphics[width=1.5in,  height=1.75in]{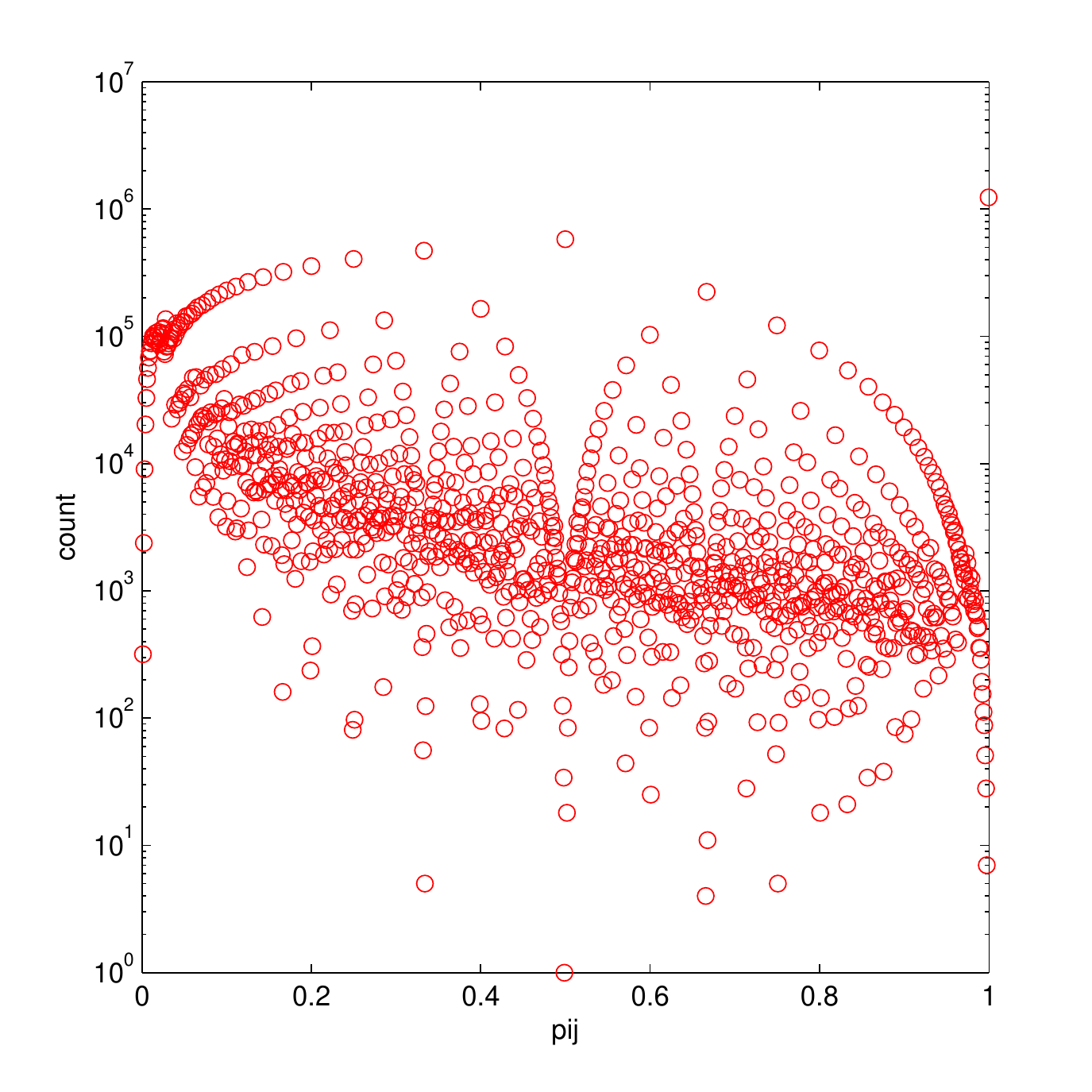}
        \label{ }
	}
\subfigure[\scriptsize  Prop. of $j \rightarrow i$ calls ]  {
        \includegraphics[width=1.5in,  height=1.75in]{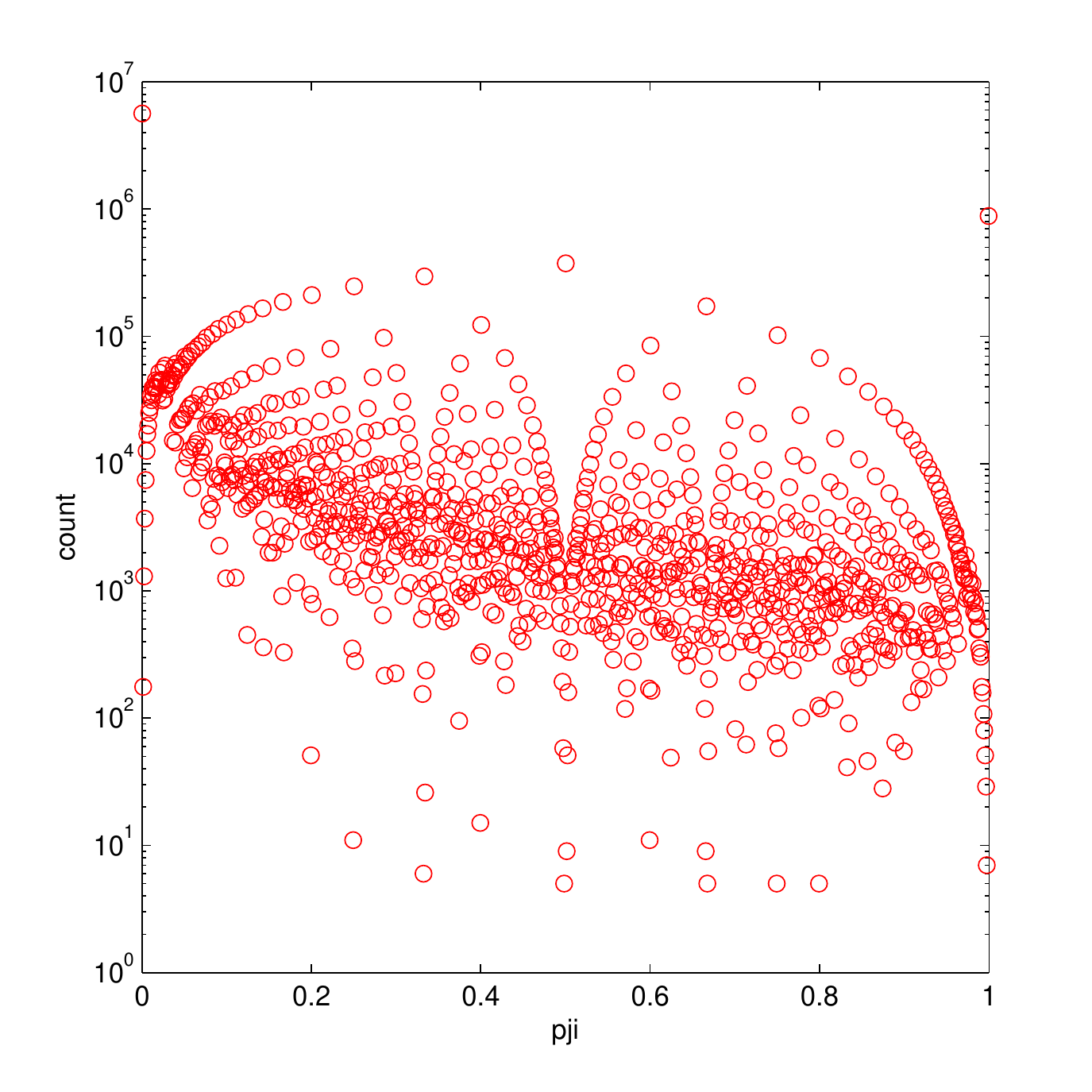}
        \label{ }
	}
\subfigure[\scriptsize  N.  of common neighbors] {
        \includegraphics[width=1.5in,  height=1.75in]{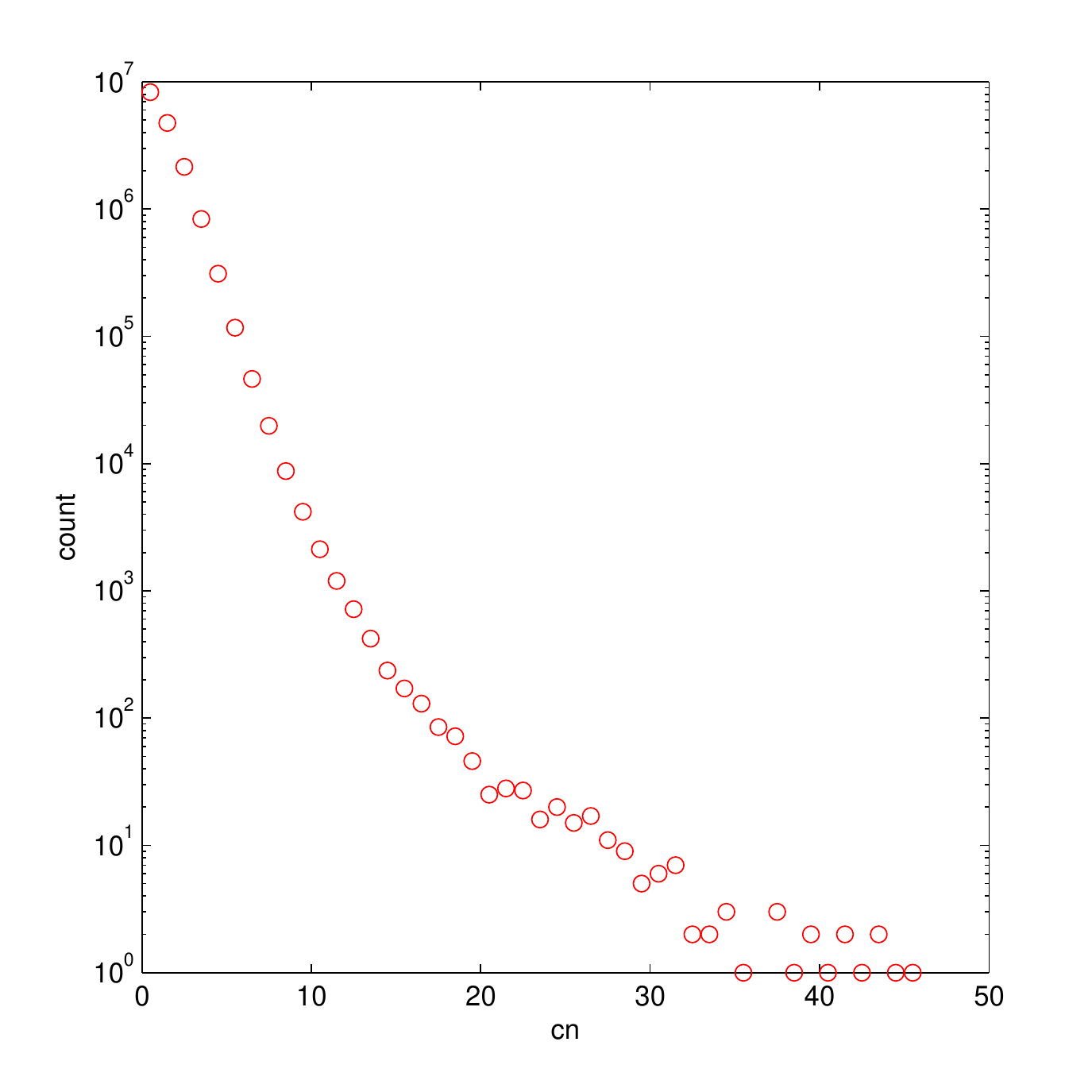}
        \label{ }
	}
\subfigure[\scriptsize  N.  of $i$ ’s neighbors that call $j$] {
        \includegraphics[width=1.5in,  height=1.75in]{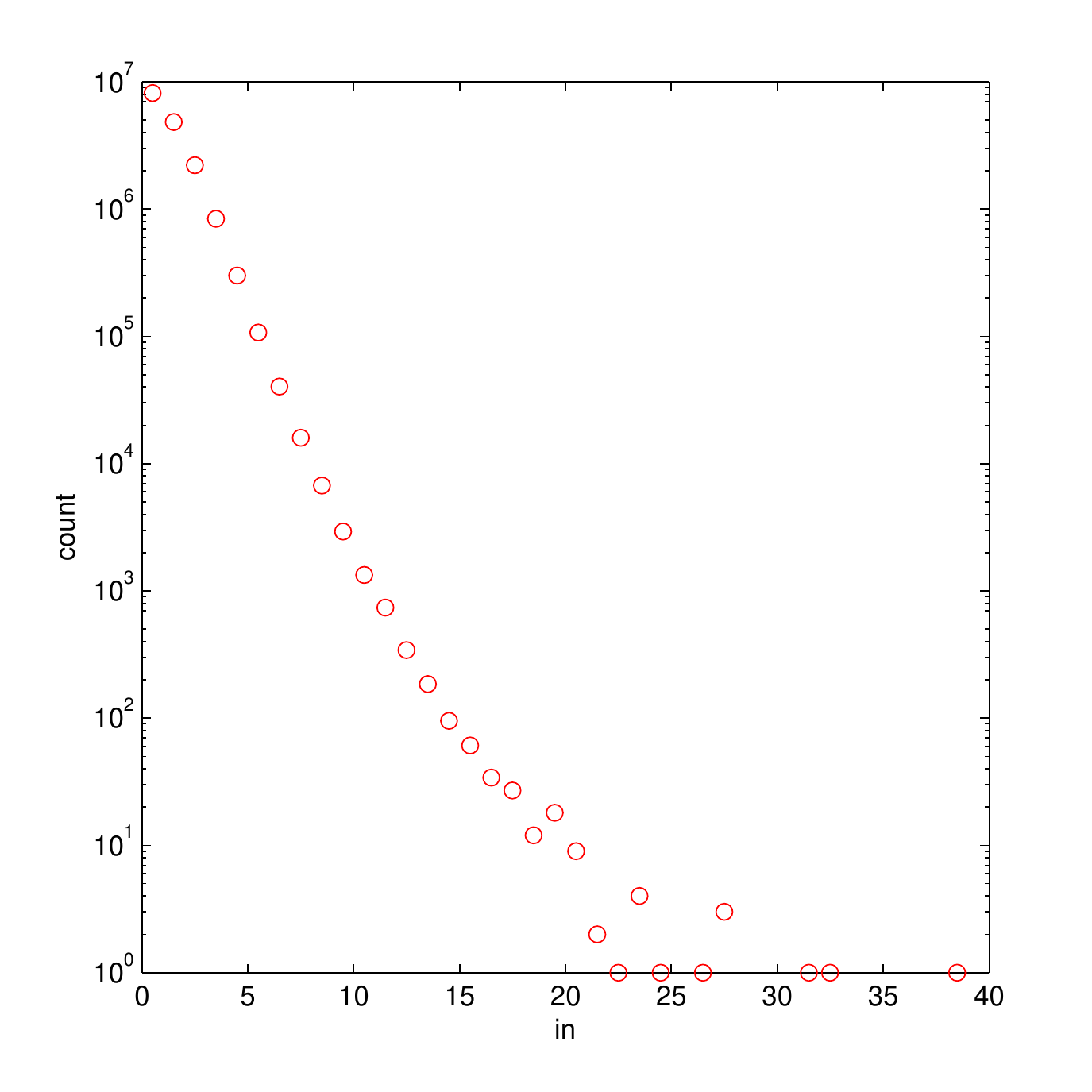}
        \label{ }
	}
\subfigure[\scriptsize N.  of $j$ ’s neighbors that call $i$ ] {
        \includegraphics[width=1.5in,  height=1.75in]{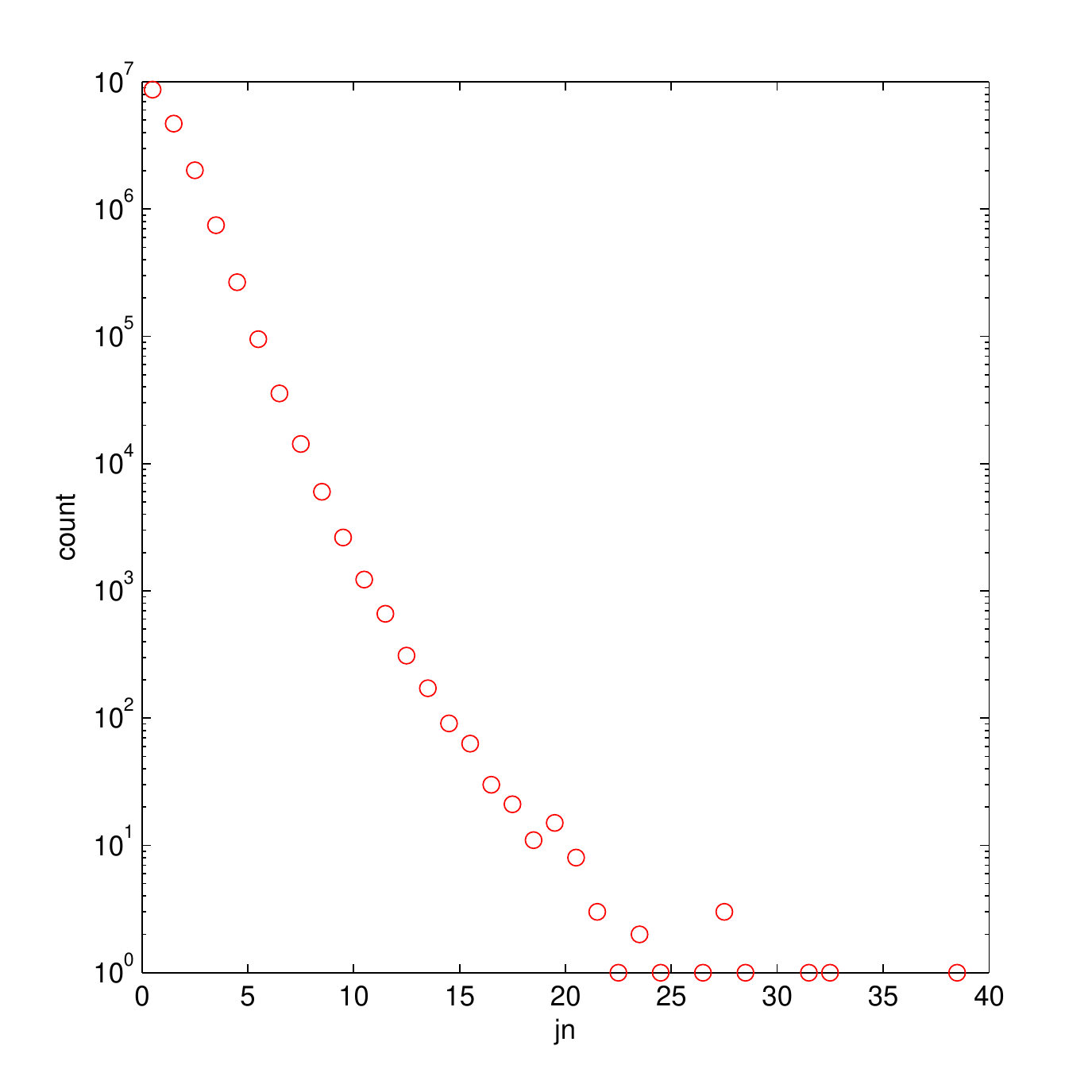}
        \label{ }
	}
\subfigure[\scriptsize N.  of calls from $i \rightarrow j$'s neighbors] {
        \includegraphics[width=1.5in,  height=1.75in]{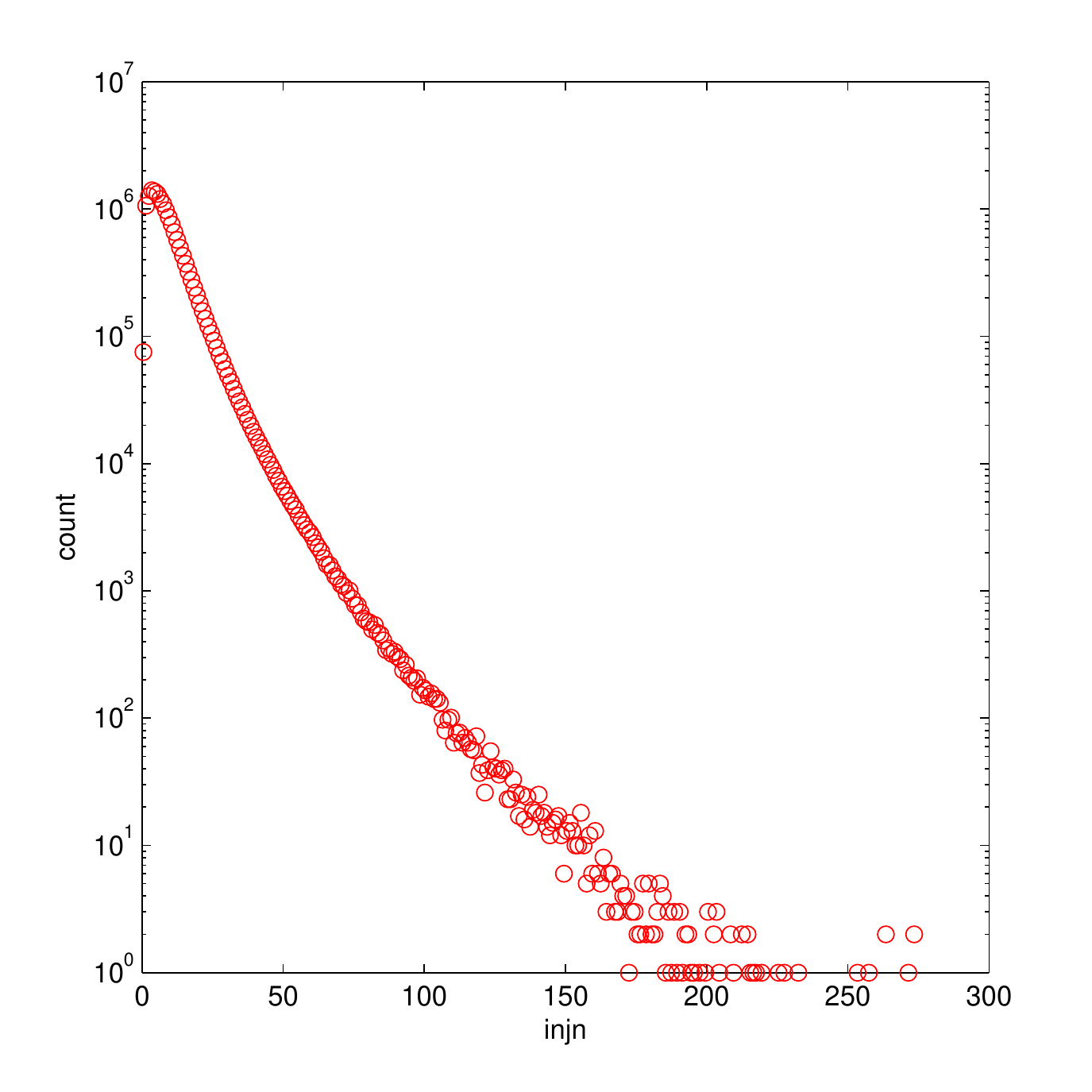}
        \label{ }
	}
\subfigure[\scriptsize N.  of calls from $j \rightarrow i$'s neighbors] {
        \includegraphics[width=1.5in,  height=1.75in]{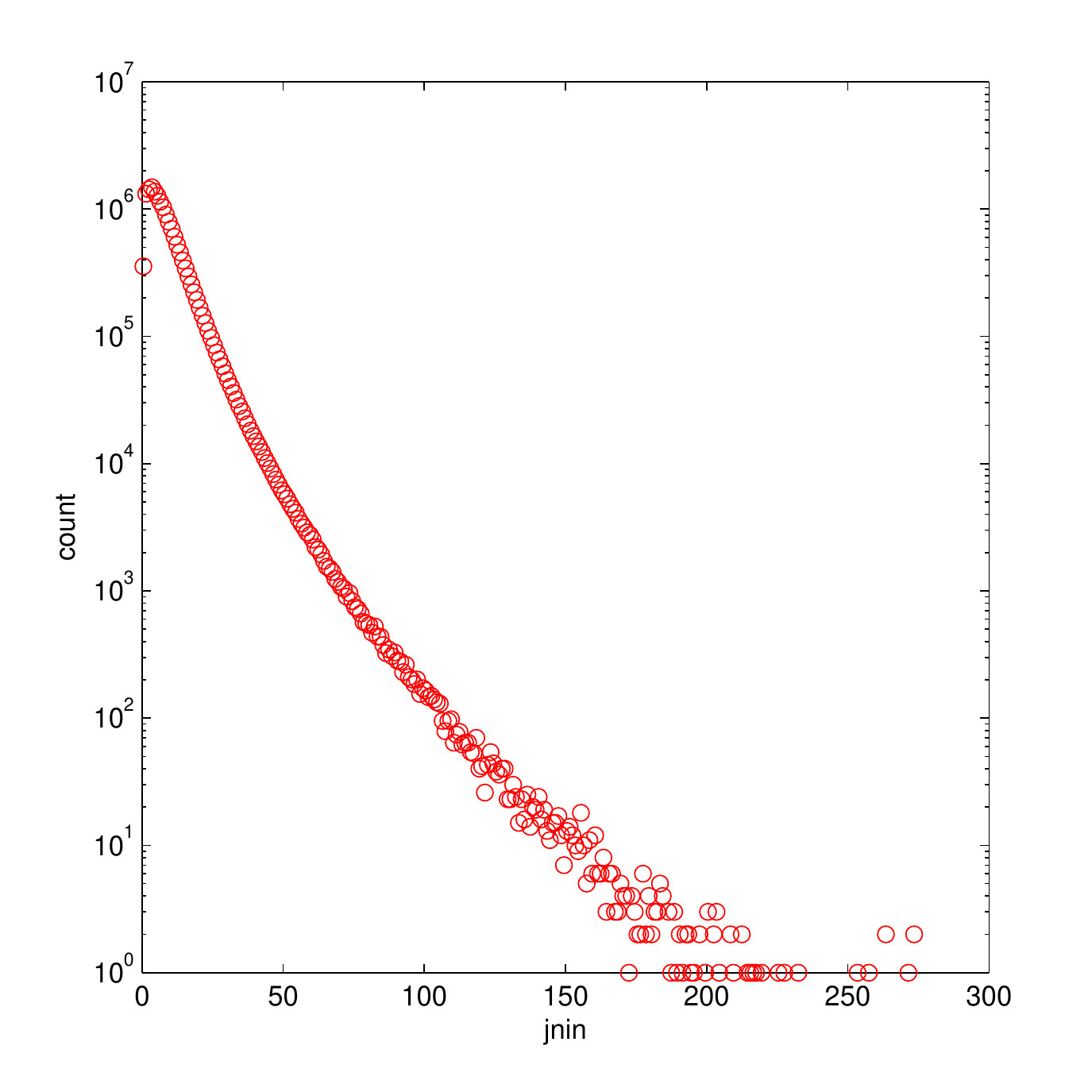}
        \label{ }
	}
\subfigure[\scriptsize Time of first call from $i$ to $j$  ] {
        \includegraphics[width=1.5in,  height=1.75in]{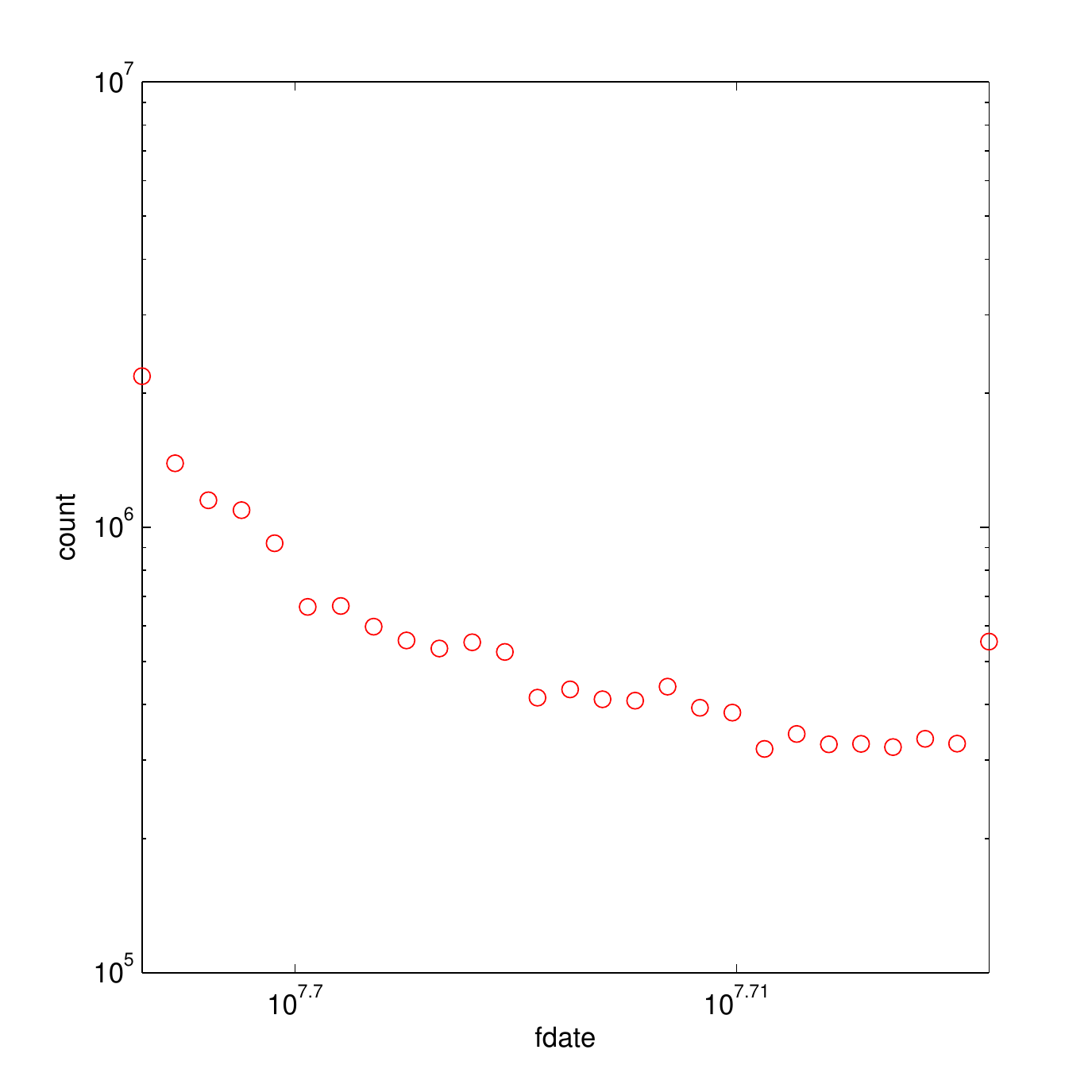}
        \label{ }
	}
\subfigure[\scriptsize  Time of last call from  $i$ to $j$ ] {
        \includegraphics[width=1.5in,  height=1.75in]{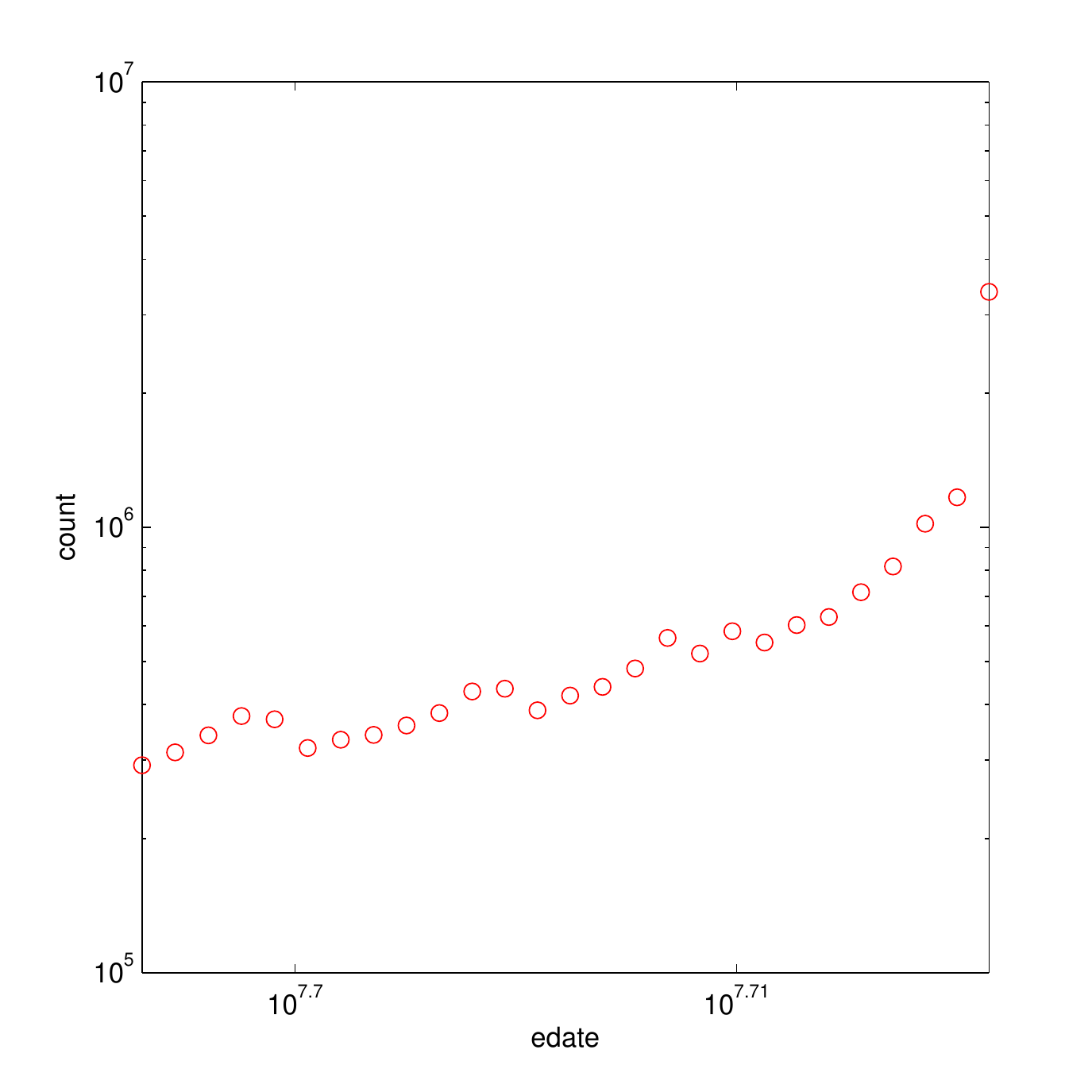}
        \label{ }
	}
\label{fig:dists}
\caption{Cumulative distributions of the features included in the analysis.}
\end{figure*}

A tree induction algorithm will then recursively divide up each of the resulting subspaces until some stopping criterion (i.e. a minimum number of instances per leaf or minimum leaf purity) is met.  Figure~\ref{fig:toy-tree} shows the decision tree generated by the splits corresponding to the ``data'' in the left-hand side.  Unknown instances are classified by taking the appropriate branches of the tree until a leaf is reached.  The class assigned to the unknown instance is whichever class was most common among the training instances at that leaf.  The Figure shows that any instance with $x > 5$ and $y > 10$ is classified as a red square, while everything else is a blue circle.  The primary advantage of decision trees for our decay prediction task (besides, of course, reasonable performance) is \textit{interpretability}.  Examining the classification accuracies at the individual leaves, we can see where the model is strong and where it is weak.  Additionally, decision trees enable us to show that the importance of the features defined in Table~\ref{tbl:features}.

\begin{table*}[ht]
	\centering
	\begin{tabular}{|l| l c c|}
		\hline
		Feature&Description&Range&Median\\ 
		\hline
		\hline
		vertex Level& & & \\ 
		\hline
			$d_{i}$&Out degree of $i$ (Ego-network Range)&1-49&2\\ 
			$d_{j}$&Out degree of $j$ (Ego-network Range)&0-49&1\\ 
			$c_{i}$&Number of calls made by $i$ (gregariousness)&1-1366&22\\ 
			$c_{j}$&Number of calls made by $j$  (gregariousness)&0-1366&22\\ 
		\hline
		Dyad Level& & & \\ 
		\hline
			$c_{ij}$&Calls from $i$ to $j$  (directed edge strength)&1-1341&2\\ 
			$c_{ji}$&Calls from $j$  to $i$ (reciprocated edge strength)&0-1341&1\\ 
			$p_{ij}$&Proportion of $i$’s calls that go to $j$  ($ c_{ij}/c_i$)&0-1&0.15\\ 
			$p_{ji}$&Proportion of $j$ ’s calls that go to $i$ ($c_{ji}/c_j$)&0-1&0.08\\ 
		\hline
		Triad Level& & & \\ 
		\hline
			$cn$&Number of common neighbors between $i$ and $j$  (edge embededness)&0-46&0\\ 
			$in$&Number of $i$’s neighbors that call $j$ (directed edge embededness)&0-39&1\\ 
			$jn$&Number of $j$ ’s neighbors that call $i$ (directed edge embededness)&0-39&0\\ 
			$injn$&Number of calls that $i$'s neighbors make to$j$ 's neighbors (2nd order embededness)&0-274&7\\ 
			$jnin$&Number of calls that $j$ 's neighbors make to $i$'s neighbors (2nd order embededness)&0-274&6\\ 
		\hline
		temporal& & & \\ 
		\hline
			$fdate$&Normalized time of first call from $i$ to $j$  (edge newness)&0-1&0.26\\ 
			$edate$&Time of last call from $i$ to $j$  (edge freshness)&0-1&0.74\\ 
		\hline
	\end{tabular}
	\caption{List of features to be used in predicting edge persistence/decay $i \rightarrow j$.}
	\label{tbl:features}
\end{table*}

\subsection{Outline of the empirical analysis}
In what follows, we consider the following three empirical issues within our chose time window ($\tau$):

\begin{enumerate}
	\item {\textit{Feature correlation}: In the initial time window $\tau_1 = (t, t + {\Delta}t)$, what is the correlation structure of the features shown in Table~\ref{tbl:features} ?} 	
	\item \textit{Feature predictiveness}: Having observed the network in a time window $\tau_1 = (t, t + {\Delta}t)$, which features of the network are most predictive of the class membership of each edge (persistent/decayed) in the adjacent time window $\tau_2 = (t + {\Delta}t, t + {\Delta}t + {\Delta}t')$? 
	\item {\textit{Edge-class Prediction}: Given a set of feature-predictors from the initial time window $\tau_1 = (t, t + {\Delta}t)$, can we build a model that accurately predicts the class membership of the edges observed in the following time window $\tau_2 = (t + {\Delta}t, t + {\Delta}t + {\Delta}t')$?}
\end{enumerate}

After briefly considering some basic descriptive statistics on each of the predictor features in the next section, in section~\ref{sec:correlation} we shed light on the first question by examining  the pairwise Spearman correlation coefficients ($\rho$) among all pairs of features in Table~\ref{tbl:features}; in section~\ref{sec:prediction} we address the second question using an information-theoretic measure of randomness for predicting short-term decay. Finally Section~\ref{sec:model} addresses the final question by formulating the edge-decay prediction task as a binary classification problem using a machine learning data analysis strategy.

\begin{figure*}[ht]
\centering
       \includegraphics[width=6in]{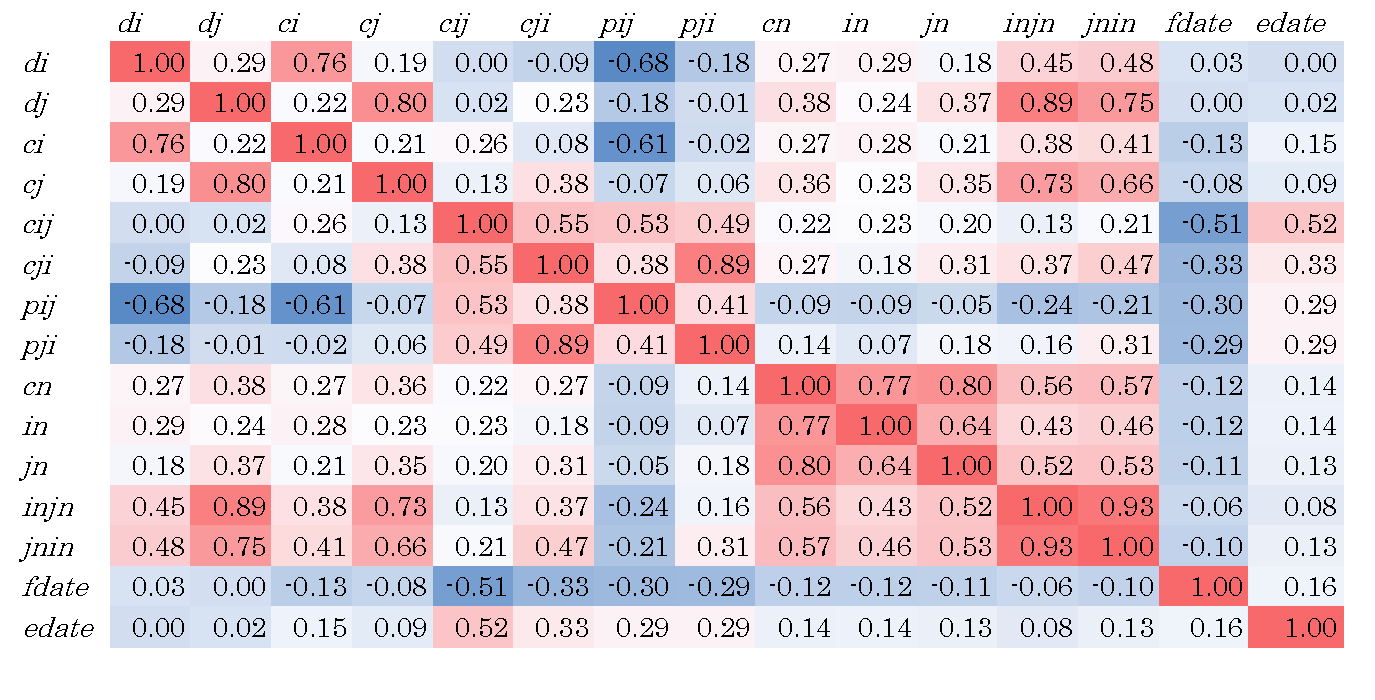}      	
	\caption{\emph{Spearman} Correlation ($\rho$) between pairs of features in one week of call data.}
	\label{fig:correl1}
\end{figure*}

\section{Feature statistics}
\label{sec:statistics}
The range and median values on the features are computed based on data from the first 4 week time period and summary statistics are provided in Table~\ref{tbl:features}.  As shown in Figure~\ref{fig:dists} of the distributions are highly skewed with substantially more lower than higher values so we report medians.  As noted earlier we omit edges with vertices which have degrees greater than or equal to 50 in order to eliminate robot calling, so vertex degree ranges from 1-49.  Note that we have included asymetric edges, that is edges in which $i$ called $j$ during $\tau_1$, but $j$ did not return a call during that time period, so there are nil values for $d_{j}$, $c_{j}$, and $c_{ji}$.  

During this $\tau_1$ period the median outdegree for the focal vertex $i$ is 2, but for its paired vertex $j$ 1 because of the presence of asymmetric edges in which $d_i^{out} >0$ and $d_j^{out} = 0$. The median value for the number of calls made by subscribers is 22.  At the dyad level, the median number of calls from $i$ to $j$ is 2, but from $j$ to $i$ it is 1, again because of asymmetric edges.  $c_{ji}$, therefore should be viewed as an indicator or reciprocity in that it indicates the extent to which vertex $j$ makes calls to $i$ given that  $i$ made at least one call to $j$.  $p_{ij}$ and $p_{ji}$ are normalized versions of $c_{ij}$ and $c_{ji}$ respectively and indicate what proportion of the total calls made by the $i^{th}$ subscriber went to each of its $j$ neighbors.  The median value of $p_{ij}$ of 0.15 indicates that for the edge at the middle of the $p_{ij}$ distribution about 15\% of its total calls went to its neighbor.  

Turning to the neighborhood-level features, the vertices joined by the median edge do not share a common neighbor as indicated by the median value of 0 for $cn$.  There appears to be a good deal more embeddedness when we measure it in terms of the edges between  $i$ and $j$'s neighbors instead of directed edges from  $i$'s  (or $j$'s)  neighbors to $j$ (or  $i$); the neighbors of vertices joined by the median edge are expected to make about seven calls to one another.  Finally for the two temporal features we have normalized them so their values range from 0-1. Values on $fdate$ and $edate$  indicate where, in the 4 week time period, the relevant events occurred. For $fdate$, the newness of the edge, the median value is .26 indicating that 50\% of the edges were active (a call had been made from $i$ to $j$) at or before about one week had elapsed. For $edate$, the freshness of an edge, the median value of .74 indicates that for 50\% of the edges the last call was made by $i$ to $j$ before three weeks had transpired.     

\begin{table*}
	\centering
	\begin{tabular}{|l | l  | c|}
		\hline
		\textbf{Feature}& \textbf{Description} &\textbf{Info. Gain}\\
		\hline
		\hline
			$d_{i}$ & Degree of $i$. &0.00235   \\
	 		$d_{j}$ & Degree of $j$. &0.00234 \\
	 		$c_{i}$& Calls made by $i$.&0.01181 \\
	 		$c_{j}$& Calls made by $j$.& 0.00449   \\
 		\hline
 		\hline
			$c_{ij}$&Calls from $i$ to $j$. &0.17948 \\
			$c_{ji}$ & Calls from $j$ to $i$. &0.12823 \\
			$p_{ij}$& Proportion of $i$'s calls that go to $j$&0.13318 \\ 		
	 		$p_{ji}$& Proportion of $j$'s calls that go to $i$.&0.12043 \\		
		\hline
		\hline
	 		$in$& Number of $i$'s neighbors that call $j$.&0.02478 \\		
	 		$jn$& Number of $j$'s neighbors that call $i$'s neighbors. &0.02303  \\
	 		$cn$& Number of common neighbors between $i$ and $j$. &0.02441 \\
	 		$jnin$& Number of $j$'s neighbors that call $i$'s neighbors. & 0.01501 \\ 		
	 		$injn$& Number of $i$'s neighbors that call $j$'s neighbors. & 0.00493  \\
 		\hline
 		\hline
	 		$fdate$& Time of first call from $i$ to $j$. &0.05104 \\
	 		$edate$& Time of last call from $i$ to $j$.&0.09954\\
		\hline
	\end{tabular}
		\caption{Information gain of each feature for predicting short-term edge decay using four weeks of data.  The information gain measures the conditional ability of that feature to predict edged decay in the subsequent week within levels of the other features.}
		\label{tbl:infogain}
\end{table*}

\section{Feature correlations}
\label{sec:correlation}

Figure~\ref{fig:correl1} shows the Spearman correlations that capture the association between features during the first four week time period.  
A number of very strong correlations (shaded red---for positive correlations---and blue---for negative correlations---in the figure) are immediately apparent.  Among the vertex level features, degree ($d_i$ or $d_j$) and gregariousness ($c_i$ or $c_j$) are highly correlated (.76 and .80) indicating that vertices with more neighbors make more calls. Among the dyadic level features the normalized and raw features of directed edge weight are correlated given that the normalized measure ($p_{ij}$) is a function of the raw measure ($c_{ij}$).  The correlation between $c_{ij}$ and $c_{ji}$ is positive indicating the presence of reciprocity, that is as the number of calls from $i$ to $j$ increases so does the number of calls from $j$ to $i$.  However this correlation is not extremely high, indicating that dyads do vary in their level of reciprocity.  Among the neighborhood-level features, the correlations are positive and for the most part large, which could indicate that the simplest measure of embeddedness, the number of common neighbors ($cn$), is a good enough measure and one does not need to look at directional or second-order embeddedness.  Finally it is interesting to note that the two temporal features, $fdate$ and $edate$, are independent.  edges that are relatively older (i.e. a call occurred earlier in the 4 week time period) can be more or less current.  

Looking at the correlations \emph{across} the various categories of the features, what is striking is how low they are.  For the most part vertex, dyadic, neighborhood and temporal features are independent from each other.  One exception to this pattern is the correlation between outdegree ($d_i$ or $d_j$) and the normalized edge weight features, $p_{ij}$ or $p_{ji}$.  This has to be the case because the sum of the $p_{ij}$'s for a give $i$ is 1, so the more neighbors a person has, the lower the proportion of their calls going to each neighbor must be (the so-called bandwidth/range trade-off \citep{aral_alstyne07}).  Another exception to the pattern of low correlations between features in different categories, is the high correlations between outdegree (and gregariousness) and the second order embeddedness features, $injn$ and $jnin$.   Agents that have more neighbors are also going to have more edges between their neighbors and the neighbors of the other vertices to which they are connected.  As such our second order embeddedness features essentially reduce to indicators of vertex  range.  Finally, the two ``temporal" features, $fdate$ and $edate$ are correlated with edge weight ($c_{ij}$ and to a less extent $c_{ji}$).  Recall that $fdate$ features the newness of an edge and that lower values are indicative of older edges.  The negative correlation of $fdate$ with $c_{ij}$ indicates, therefore, that newer edges are weaker and older edges are stronger.  The positive correlation of $edate$ with $c_{ij}$ indicates that fresher edges (i.e. edges in which a call has been made more recently) are also stronger. 

In sum, it appears that there are really four relatively independent sets of edge features pertaining to vertex, dyadic, neighborhood and temporal levels.   While there are multiple indicators within each of these sets, they tend to be highly correlated, with the exception of the two "temporal" features.  Though in the remainder of the paper we will be looking at the predictive value of all these features, based on these correlations our focus will be on the following potentially important features: outdegree (both $d_i$ and $d_j$), edge weight ($c_{ij}$), reciprocated edge weight ($c_{ji}$), the number of common neighbors ($cn$), and both the newness of the edge ($fdate$) and its freshness ($edate$).   

\section{Feature Predictiveness}
\label{sec:prediction}

We wish to determine the extent to which each of the above features as observed in the first time window helps us classifies edges as either decayed or persistent in the following time-window.  By determining this, we can quantify, to some extent, the usefulness of the features for decay prediction.  There are several possible indicators of predictive ability.  Here we rely on the \textit{information gain}, which is the standard measure of feature predictiveness in data-mining \citep{witten_frank05}.  Approaches to determine the importance of predictors based on information theory are common in statistics \citep{menard04, gilula_haberman01}.  Information-theoretic approaches have been applied before in the characterization of overall structural features of social networks (e.g. \citep{butts01, leydesdorff91}; here they are deployed in the interest of quantifying the predictive ability of fine-grained (local) structural features for the link prediction problem.

\subsection{Formal definition of Information Gain}
Information gain tracks the decrease in \textit{entropy} associated with conditioning on an attribute, where entropy is a measure of the randomness (alternately, predictability) of a quantity.  To understand the measure and how it quantifies feature importance, consider as an example our two-class  cell phone dataset, where each edge is either persistent (class 1) or decayed (class 0).  If we define $p(x)$ as the proportion of instances of class 1 and $q(x)$ as the proportion of class zero, the entropy $H(x)$ is defined as: 

\begin{equation}
H(x) = -p(x) \ log \ p(x) - q(x) \ log \ q(x)
\end{equation}

\noindent where all logarithms are taken to base 2.  If the two classes are perfectly balanced, then the entropy $H(x) = log \ 2 = 1$.  As the classes become increasingly imbalanced, the entropy decreases.  That is to say, we know more a priori about the class of a random instance.  If a particular feature is informative, then conditioning on that feature should decrease the entropy of the dataset.  Suppose, for example, that a feature $F$ takes a set $K$ of possible values.  The \textit{conditional entropy} of the dataset conditioned on the feature $F$ is:

\begin{equation}
H(x | F) = \sum_{k \in K}{-p_k(x) \ log \ p_k(x) - q_k(x) \ log \ q_k(x)}
\end{equation}

Where $p_k(x) = p(class = 1 | F = k)$ is the proportion of positive-class instances among the instances where the feature $F$ takes the value $k$.  Similarly, $q_k(x)$ is the proportion of negative-class instances.  The \textit{information gain} for the feature $F$ is the decrease in entropy achieved by conditioning on $F$: $I(x | F) = H(x) - \frac{H(x | F)}{|K|}.$

Returning to our hypothetical example, suppose there is a feature $F$ that  takes on two values: $A$ and $B$.  Instances with $F = A$ are 90\% class 1 and instances with $F = B$ are 90\% class 0, then $I(F) = log \  2 - (-\frac{9}{10} log \ \frac{9}{10} - \frac{1}{10} log \ \frac{1}{10}) = 0.530.$  The information gain $I(x | F)$ has an appealing intuitive interpretation as the percentage of information about the class that is revealed by the feature $F$.  By  calculating the information gain of each feature in Table~\ref{tbl:features}, we can determine which actor and edge attributes reveal the most information about edge decay. 

\subsection{Feature Importance in the Call Network}

Table~\ref{tbl:infogain} shows the information gain of each feature described in Table~\ref{tbl:features} calculated for the first four weeks of data.  The results show that the four most predictive features are \emph{dyadic} features of directed \emph{tie strength} as given by the frequency of interaction and the extent to which communications are concentrated on a given alter: number of calls sent and received along the edge ($c_{ij}$ and $c_{ji}$) and call proportions from both $i$ to $j$ and $j$ to $i$.  There is a substantial drop-off in the information gain produced by the remaining features. After the dyadic-level features, the most important predictors are associated with the observed age of the tie and the recency of communication ($fdate$ and $edate$, respectively).  Here we observe that time of first call between $i$ and $j$ (edge newness)  is only about half as predictive as the time of last call between $i$ and $j$ (edge freshness) ($(I(decay|fdate)=0.05$ versus $(I(decay|edate)=0.10$), suggesting that freshness beats newness as a predictive criterion.  These are followed, in terms of predictive ability, by the neighborhood-level (e.g. number of common neighbors and frequency of interaction among neighbors of the two members of the dyad) and the vertex-level features.  The predictiveness afforded by either vertex or neighborhood level features is comparatively minimal.   

These results suggest that previous research on tie decay, which has for the most part been unable to consider the strength of individual ties (as it has limited itself to binary network data), may have missed the most critical single factor for tie decay.  This raises the question: Do features that have previously been deemed important (such as embeddedness, newness, and range actually drive tie decay or are they merely correlated surrogates for tie strength and therefore appear important only when concrete measures of strength are absent? 

\begin{table*}[ht!]
	\centering
	\begin{tabular}{|l | c c | c c|}
		\hline
		&\multicolumn{2}{|c|}{Tree}&\multicolumn{2}{|c|}{Logistic}\\
		\hline
		\hline
		&Persist&Decay&Persists&Decay\\
		\hline
		Accuracy&0.737&0.737&0.734&0.734\\
		Precision&0.780&0.684&0.796&0.668\\
		Recall&0.754&0.714&0.722&0.751\\
            \emph{F}&0.767&0.699&0.757&0.707\\
            \hline
	\end{tabular}
	\caption{Comparison of model fit-statistics for the decision-tree and logistic regression classifiers. }
	\label{tbl:classperf}
\end{table*}

\section{Predicting edge persistence and decay}
\label{sec:model}
\subsection{Classifier comparison}

Table~\ref{tbl:classperf} summarizes the performance of our two classifiers under all four prediction scenarios.  It presents four standard performance metrics: \textit{accuracy}, \textit{precision}, \textit{recall}, and \textit{F-Measure}.  \emph{Accuracy} is the proportion of all instances that the model correctly classifies.  The other three metrics measure the types of error made by the classifier.  \textit{Recall} gives the proportion of observed persistent ties that the model correctly classifies as persisting while \textit{precision} gives the proportion of ties that the model predicted as belonging to the persistent class that actually did persist.  Precision and recall, to some extent, measure two competing principles.  Theoretically, a model could achieve very high recall by classifying all ties as persistent, but such a model would have very low precision.  Similarly, a model could achieve perfect precision by classifying only its most confident instance as positive, but in doing so, it would achieve very low recall.  The \textit{F-Measure} captures the trade-off between precision and recall.  This is defined as the harmonic mean of precision ($P$) and recall  ($R$):

\begin{equation}
F = \frac{2PR}{P + R}
\end{equation}

\noindent where $P$ is the precision and $R$ is the recall of the model in question.

We evaluate both classifiers on both the majority class, persistence (57\% of dyads), and the minority class, decay (43\%).  A classifier is expected to do better on the majority class because there is more available data with which to build the prediction.   As shown in the first two columns of Table~\ref{tbl:classperf}, the decision-tree classifier performs reasonably well in regards to the majority class: it correctly predicts 73.7\% of all ties (accuracy) and 75.4\% of all persisting ties (recall).  In regards to the minority class, decay, the decision-tree classifier does a little bit worse. The decision-tree classifier correctly predicts 71.4\% of all decaying ties.  The model is also less precise when it comes to predicting decay.  About 68.4\% of ties predicted to decay do in fact decay, while in the case of persistence about 78\% of the ties that the model predicts persist do in fact persist.  Overall the decision tree classifier does a good job and shows tie persistence in social networks is fairly predictable in the short-term from local structural, temporal and vertex-level information.

The last two columns of Table~\ref{tbl:classperf} present these same fit statistics when we use the logistic regression classifier for the decay/persistence prediction task.  The results are very similar to the results obtained when using the C4.5 decision tree model.  The logistic regression  correctly predicts 73.4\% of all ties (accuracy), and for the majority class about 72\% of persisting ties are correctly classified by the regression model (recall).  In contrast to the decision tree model, the recall values are higher for the decay class.  The logistic regression model correctly classifies about 75\% of decayed ties, while the decision tree model correctly classifies about 71.4\% of decayed ties.  This is not a big difference, but it does seem to indicate that in this case the logistic regression model does a slightly better job predicting decay, while the decision tree model does a slightly better job  predicting persistence.  However, the precision results on the decay class are slightly worse in the logistic regression model compared to the decision tree model.  The result is that the $F$-statistic is about the same across the two models.

In sum, both the decision tree and logistic regression classifiers indicate that tie persistence and decay patterns are predictable and that using either model yields fairly similar levels of prediction and error.  The consistency between these two ways of modeling the data---a more standard regression approach and a relatively non-standard data mining approach---gives us confidence in the results.  After presenting the results of the logistic regression coefficients in the next section, we turn to the decision tree results and show how they yield new insights about what is predicting tie persistence/decay in social networks.

\begin{table*}[ht!]
	\centering
	\begin{tabular}{|l  l|c c|}
		\hline
		Feature&Description&$\beta$&Odds ($exp(\beta)$)\\ 
		\hline
		\hline
			$d_{i}$ & Degree of $i$ &-0.0335&0.9671\\ 
			$d_{j}$ & Degree of $j$ &0.0057&1.0057\\ 
			$c_{i}$& Calls made by $i$ &0.0003&1.0003\\ 
			$c_{j}$& Calls made by $j$ &-0.0013&0.9987\\ 
		\hline
		\hline
			$c_{ij}$&Calls from $i$ to $j$& 0.0373&1.0380\\ 
			$c_{ji}$ & Calls from $j$ to $i$&0.0229&1.0232\\ 
			$p_{ij}$& Proportion of $i$'s calls that go to $j$ &0.0504&1.05178\\ 
			$p_{ji}$& Proportion of $j$'s calls that go to $i$ &0.8521&2.3446\\ 
		\hline
		\hline
			$in$& Number of $i$'s neighbors that call $j$. &0.1409&1.1513\\ 
			$jn$& Number of $j$'s neighbors that call $i$'s neighbors. &0.0877&1.0917\\ 
			$cn$& Number of common neighbors between $i$ and $j$.& 0.0525&1.05391\\ 
			$jnin$& Number of $j$'s neighbors that call $i$'s neighbors. &-0.0366&0.9641\\ 
			$injn$& Number of $i$'s neighbors that call $j$'s neighbors. &0.0416&1.0425\\ 
		\hline
		\hline
			$fdate$& Time of first call from $i$ to $j$.&-2.3021&0.1000\\ 
			$edate$& Time of last call from $i$ to $j$. &2.9218&18.5747\\ 			
		\hline
	\end{tabular}
	\caption{Logistic regression coefficients of the effect of each feature in predicting edge-persistence.}
	\label{tbl:logistic}
\end{table*}

\subsection{Logistic regression classifier results}
\label{subsec:logit}

Table~\ref{tbl:logistic} shows the parameter estimates from the logistic regression model (predicting the log-odds of a tie persisting) along with the odds-ratios.  The estimates are based on a full model including all the features.  We do not report standard errors as all the estimates are statistically significant given the large size of the training data on which these parameters are estimated.   

Beginning with the features that our information gain values indicated were likely to be the most important (see table~\ref{tbl:infogain} and the discussion in section~\ref{sec:prediction}) we see that the call volume from $i$to $j$  (directed tie strength, $c_{ij}$) has a positive effect on persistence.  For each additional  call made, the odds of the tie persisting is almost 4\% higher. Net of this influence, the number of calls that $j$ makes back to $i$,  $c_{ji}$, is also positive.  For each additional reciprocating call, the odds of a tie persisting increases about 2\%.

The effects of the outdegree of each member of the dyad have opposite signs. In general, an edge that starts from a vertex with a large number of neighbors has a higher chance of decaying. However, if that edge is directed at a vertex of high-degree, then it has higher chances of persisting.  These effects have a straightforward interpretation, high-degree actors have less persistent edges, but this effect is mitigated when these edges are directed towards other high-degree actors.\footnote{This suggests that bulk of the fluctuating, low-persistence edges characteristic of high-degree actors are those which are directed towards actors of low degree.  When popular actors connect to other popular actors, their relationships tend to be more stable than when they connect to low-degree alters. Conversely, while low-degree actors tend to have---on average----more stable relationships, these become even more stable when directed at more popular alters.} The other two vertex-level features pertaining to gregariousness ($c_i$ and $c_j$) have very small effects.  This indicates that after adjusting for degree, raw communicative activity does not appear to be involved in processes of edge persistence and decay.

Turning to the neighborhood-level measures, all the effects are positive except for the 2nd order embeddedness measure $injn$, which as noted earlier is  correlated with $d_i$ and $d_j$.  In general embeddedness increases the odds that a tie will persist, consistent with previous research that show that embedded edges decay at a slower rate \citep{burt97, burt00}.  For each additional common neighbor between i and j, the odds of a tie persisting increases 5.4\%.  The directed embeddedness measures $in$ and $jn$ appear to be even stronger.  For example, for each additional neighbor of i that calls j the odds of the tie persisting increases 15\%.

Finally the temporal measures have opposite effects.  $fdate$ has a negative effect on persistence indicating the newer ties (which have higher values on $fdate$) are more likely to decay, indicative of the liability of newness that \citet{burt97} notes is an important characteristics of social ties.  On the other hand $edate$, the freshness of the tie, has a positive effect on persistence.  Ties that have been activated recently are more likely to persist than those that have been inactive.  

\begin{figure*}[ht!]
\centering
       \includegraphics[height=6in]{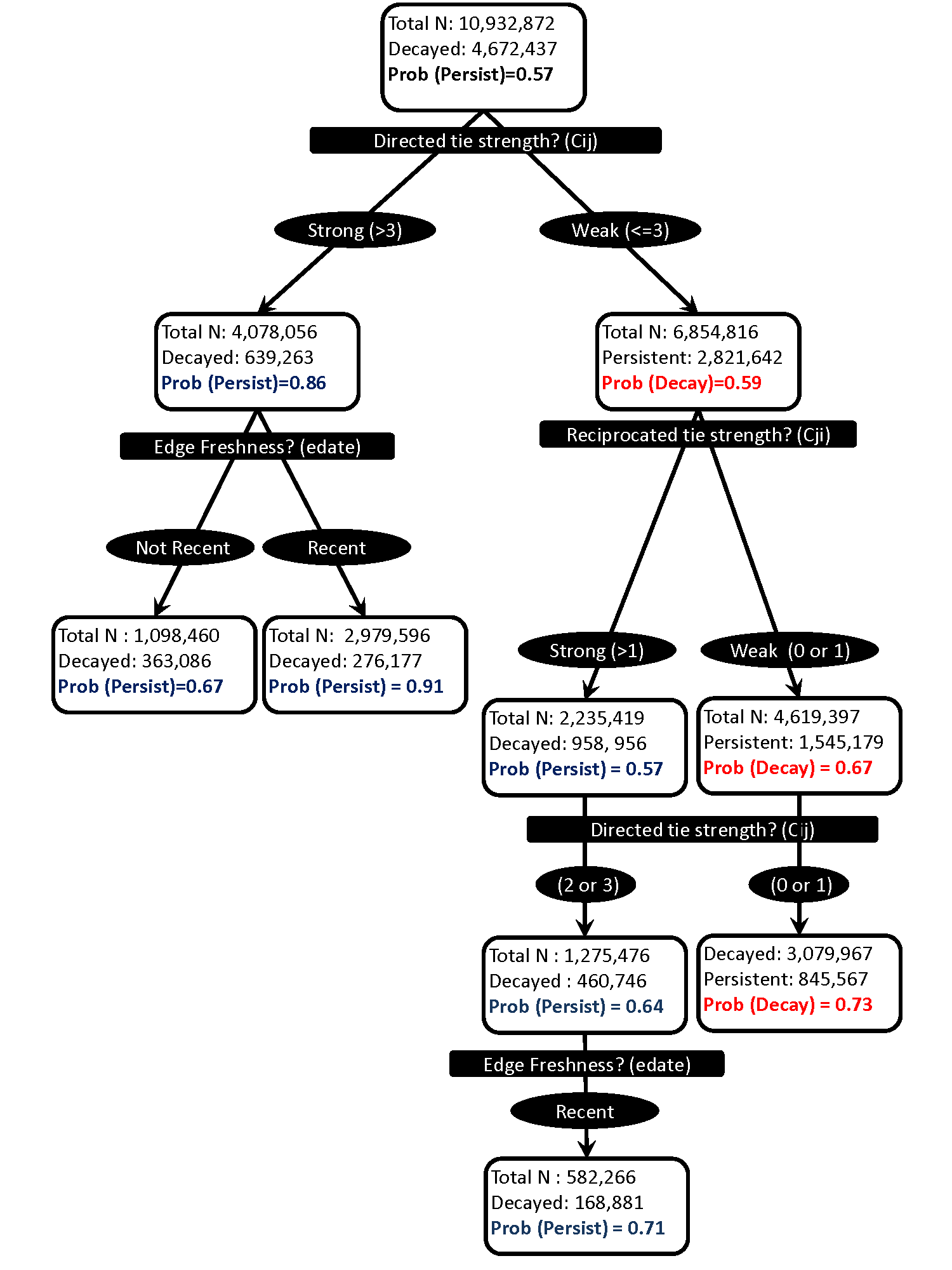}      	
	\caption{Selected leaves of the best-fitting decision-tree obtained from the training set.}
	\label{fig:pretty-tree}
\end{figure*}

\subsection{Decision-tree classifier results}
As we mentioned in Section~\ref{subsec:tree}, the  structure of decision trees can offer insights into the underlying characteristics of the data on which they were trained.  Recall that, at each subtree, our C4.5 implementation chooses the attribute with the largest information gain on the data within that subtree.  This means that, at each step, the attribute providing the greatest amount of additional information is chosen for further splitting. Figure~\ref{fig:pretty-tree} shows selected branches of the resulting decision-tree obtained from the training data.  In the figure, directed edge weight ($c_{ij}$)---as measured by the number of calls directed from one person to another ---is the strongest discriminator of class membership as we saw earlier (Table~\ref{tbl:infogain}) and thus stands as the top node of the tree.  As deeper levels of the tree we find that conditional on directed edge weight other dyadic and one temporal feature helps to predict tie decay, but not vertex-level factors such as degree and neighborhood level factors such as the number of common neighbors.

The left-hand side of the figure shows that the optimal directed edge-weight ($c_{ij}$) cutoff differentiating persistent from decayed dyads in our data is approximately 3. Dyads in which one of the actors contacted the other more than three times in the initial 4-week period  have very strong odds of being classified as active in the following  4-week period ($p=0.86$).  If in addition to that (as we follow the tree into the third level), the edge has been activated recently (has high freshness) then we can be virtually certain that they tie will persist ($p=.91$).  If the edge has not been refreshed recently, however, then the probability of persistence drops substantially  ($p=.67$)

The right hand side of the figure shows that for edges with relatively weak directed weight, the odds of decay are relatively high ($p=0.59$).  If in addition, the edge is non-reciprocal (with incoming directed strength being even weaker or equal to zero) then the probability of decay rises concomitantly ($p=0.67$).  However, even with low levels of directed strength ($c_{ij}\leq3$), an edge characterized by reciprocity has a relatively decent chance of persisting in the next period ($p=0.57$), if in addition to this the edge is on the ``high-side'' of the corresponding weight cutoff  ($2 \geq c_{ij} \leq3$), and it was also active later in the time period (has high-freshness), then the probability then the probability of being classified in the persistent class improve substantially ($p=0.71$).

\section{Discussion and Conclusion}
\label{sec:conclusions}
In this paper we explore the question of short-term decay of cell-phone contacts as a problem of \textit{decay/persistence prediction}: determining what local structural features allow us to best determine whether certain dyads that are considered to be \emph{connected} during a given time window will be \emph{disconnected} during an immediately adjacent time window.  Using large-scale data on millions of dyads from a large non-U.S. cell phone provider, we investigate to what extent we can gain empirical leverage on the decay prediction problem.  Our analytic framework is guided by prior literature on the structural and vertex-level predictors of edge-decay in informal social networks. Using observational data from call logs, we calculate features of ego-network range, communicative range, edge-strength, reciprocity, embeddedness, edge-newness and edge-freshness.  

In all we took into account a total of 15 vertex-level, dyadic, neighborhood-level and temporal features (e.g., edge weight, embeddedness, ego-network range, and newness) most of which incorporated information on the relative frequency of interaction, and thus on the \emph{ weight} associated with each component arc in the cell-phone network \citep{barrat_etal04}.  The results support our emphasis on the importance of edge weights, as we find that, according to the information gain metric (an information-theoretic measure of predictiveness) factors related to \textit{directed edge weight}---essentially the measure of total directed communicative flow within the dyad---are more predictive of decay than any of the other types of factors.   Our analysis of the correlation structure of the other types of features (vertex, dyad, neighborhood-level and temporal) with empirical indicators of edge weight suggested that while there is a reasonable amount of correlation between edge weight and these other features, it is not strong enough to conclude that edge weight is a redundant by-product of other local-structural factors.  To explore the conjoined effect of the various features on edge-decay we built a decision-tree and logistic regression classifier and evaluated their joint effectiveness at predicting short term decay in the cell-phone contact network.  We found that that both classifiers performs reasonably well.

The logistic regression classifier results are consistent with what we know about the structural and temporal dynamics of relationship persistence and decay.  Stronger ties are more likely to persist and reciprocation increases persistence as well.  While the overall calling activity of each of the actors involved in the dyad is not that important, the number of neighbors that they are connected to is, with decay increasing for outgoing ties originating from high-degree actors, but with this effect being contingent on the number of neighbors of the target actor.  This result implies that relative inequalities in network range can tell us something about the expected stability of edges in social networks, as the bulk of the ``instability'' in edge evolution may be accounted for by the activity of high-degree actors. This result is consistent to that obtained in a network constructed using email trace logs \citep{aral_alstyne07, kossinets_watts06}.   Embeddedness is also important.  When a tie is embeddded in triadic or larger structures, they are protected from fast decay.  Finally,  new ties are more likely to decay, while ties that have been active recently are more likely to persist.  

Finally, we show that the structure of the decision-tree classifier can provide useful insights on the relative importance of different vertex-level and dyadic level processes in determining the probability that particular types of edges in the cellphone network (e.g. high versus low weight) will decay. The results of the decision-tree classifier are consistent with the initial feature predictiveness results, giving us what combinations of the high-information gain features shown in Table~\ref{tbl:infogain} generate persistence and decay.  As the decision tree shows, the most important predictors are directed edge strength, reciprocated edge strength and the freshness of the tie. So while network range,  embeddedness, and tie age can be used to predict persistence as the logistic regression estimates and information gain statistics indicate, they are not the most important factors.   
 
Phrased in terms of ``rules,'' we can say that persistent edges in the cell-phone network are those characterized by high-levels of interaction frequency coupled with relatively constant re-activations (freshness) of the edge over time. Edges at high risk of decay on the other hand, are characterized by relatively low levels of interaction and nonreciprocity. Finally, a second path towards persistence appears to be characteristic of ``nascent'' edges which have yet not had the opportunity to gain strength: here relatively weak flows are combined with reciprocity and recent activation to produce persistence in calling behavior, at least in the short term. 
 
In terms of contemporary models of relationship evolution, this last result suggests that in order to persist, social relationships must first cross a boundary where the the directed attachment between ego and alter becomes ``synchronized.''  This implies that the observed strength of older relationships may be an \emph{outcome} of the achievement of reciprocity at the early stages; thus as \citet[241]{friedkin90} notes  ``{\ldots}reciprocation  and  balance  are  crucial  for  both  the  occurrence  and durability  of  a strong  relationship.''  In this respect, while strong weight---and thus frequency of interaction \citep{homans50}---is sufficient to guarantee a persistent (if in some cases asymmetric), relationship after a certain relationship-age threshold is crossed, reciprocity appears to be more important for the longer-term survival of weaker edges, especially in the nascent stages of the relationship \citep{friedkin90}.  
  
These time-dependent balance/strength dynamics therefore seems to us to deserve detailed consideration in future modeling efforts.  In this paper we have attempted establish the beginnings of a framework with which to rank factors that differentiate those links fated for quick dissolution from those that will become a more permanent component of the social structure.




\section*{References}

\bibliographystyle{elsarticle-harv}
%%\bibliographystyle{elsarticle-harv}
%%\bibliography{N:/Private/bib/main}




%% \bibitem must have one of the following forms:
%%   \bibitem[Jones et al.(1990)]{key}...
%%   \bibitem[Jones et al.(1990)Jones, Baker, and Williams]{key}...
%%   \bibitem[Jones et al., 1990]{key}...
%%   \bibitem[\protect\citeauthoryear{Jones, Baker, and Williams}{Jones
%%       et al.}{1990}]{key}...
%%   \bibitem[\protect\citeauthoryear{Jones et al.}{1990}]{key}...
%%   \bibitem[\protect\astroncite{Jones et al.}{1990}]{key}...
%%   \bibitem[\protect\citename{Jones et al., }1990]{key}...
%%   \harvarditem[Jones et al.]{Jones, Baker, and Williams}{1990}{key}...
%%


% \end{thebibliography}

\end{document}